\documentclass[jmp,aip,reprint]{revtex4-2}
\usepackage{amsmath}
\usepackage{amssymb}
\usepackage{amsthm}
\usepackage{graphicx}

\newtheorem{theorem}{Theorem}[section]
\newtheorem{lemma}[theorem]{Lemma}
\newtheorem{proposition}[theorem]{Proposition}
\newtheorem{corollary}[theorem]{Corollary}

\newtheorem{remark}[theorem]{Remark}

\begin{document}

\title{Finite-volume and thermodynamic-limit lower bounds for the coherently
driven Bose--Hubbard model via the Lieb--Yamazaki method}
\author{Maciej Janowicz}
\affiliation{Institut für Physik, Carl von Ossietzky Universität, Oldenburg}
\affiliation{Faculty of Applied Informatics and Mathematics,
Warsaw University of Life Sciences--SGGW}
\author{Joanna Kaleta}
\affiliation{Faculty of Applied Informatics and Mathematics,
Warsaw University of Life Sciences--SGGW}
\author{\begingroup\setlength{\fboxsep}{2pt}\setlength{\fboxrule}{0.6pt}%
\fbox{\strut Filip Krzyżewski}\endgroup}
\affiliation{Institute of Physics, Polish Academy of Sciences}
\author{Andrzej Zembrzuski}
\affiliation{Faculty of Applied Informatics and Mathematics,
Warsaw University of Life Sciences--SGGW}
\date{}

\begin{abstract}
We apply the Lieb--Yamazaki positive-operator decomposition to the coherently
driven Bose--Hubbard model on a periodic hypercubic lattice in arbitrary
dimension. Decomposing the Hamiltonian into a positive semidefinite part and
a quadratic remainder, we diagonalize the latter via a uniform displacement
and a Bogoliubov transformation. This yields a variational family of rigorous
lower bounds on the ground-state energy, valid within an explicit stability
region. Taking the thermodynamic limit, we establish the existence of the
lower-bound energy density and the continuous excitation band of the bounding
Hamiltonian. In the corresponding Gaussian ground state, we derive explicit
expressions for the order parameter, filling fraction, energy densities,
particle-number variance, and second-order correlation function, all of which
possess well-defined thermodynamic limits. Within this auxiliary Gaussian
state, we show that a sufficiently strong coherent drive induces antibunching
and sub-Poissonian statistics. Analyzing
the long-wavelength excitation spectrum reveals a finite energy gap and
effective mass in the stable region, while the edge of the admissible region
features a gapless spectrum with a linear boundary slope generated by pairing
terms. This slope characterizes the bounding Hamiltonian rather than the full
driven model. Finally, we provide extensive numerical comparisons between
finite-volume and thermodynamic-limit bounds to map the tightness of the
estimates and assess the quantitative performance of the Gaussian ansatz.
\end{abstract}

\maketitle

\section{Introduction}

The Bose--Hubbard model provides a standard setting for the competition
between tunneling, local interactions, and particle-number fluctuations in
lattice boson systems~\cite{Fisher1989,Jaksch1998,Greiner2002}.  Its
coherently driven form is relevant to nonlinear
bosonic arrays and to protocols in which an external field controls the
amplitude and phase of the lattice modes.  The drive breaks particle-number
conservation and introduces a linear source term, enriching the ground-state
structure while removing several simplifications available in the undriven
model.  These features make the model pertinent to strongly correlated
bosons and cold-atom physics, as well as to driven photonic lattices and
quantum fluids of light~\cite{CarusottoCiuti2013,Hartmann2016,
NohAngelakis2017,Bloch2022}.  Experiments on open photonic systems have
observed dissipative critical behavior and stabilized Mott states
~\cite{Fink2018,Ma2019}, while driven-dissipative theory has addressed
critical slowing down, excitation and correlation structure, and chiral or
chaotic dynamics~\cite{Vicentini2018,Grass2019,Dahan2022,Rassaert2025}.
Open photonic realizations motivate the coherent source term, but the present
work addresses the closed Hamiltonian sector only; dissipation and
steady-state Liouvillian spectra are outside its scope.

Variational trial states readily furnish upper bounds on the ground-state
energy, but rigorous lower bounds are considerably less accessible in the
presence of both a quartic on-site interaction and a coherent source.  A
useful lower estimate must control the unbounded local occupation, remain
stable under the thermodynamic limit, and retain explicit dependence on the
detuning, hopping, and drive.  It is also desirable to determine what the
bounding construction implies for observables and excitations, rather than
for the energy alone.  Rigorous Bogoliubov analyses and energy estimates for
interacting Bose gases provide complementary control of quadratic
approximations and their remainders~\cite{Lewin2015,Napiorkowski2018,
Brietzke2020}. Elementary one-site and bond-cluster estimates also follow
from operator inequalities and Anderson's cluster lower-bound construction
~\cite{Anderson1951}. We therefore compare the Lieb--Yamazaki result directly
with these important baselines rather than claiming uniqueness of the bound.

Herein, the Lieb--Yamazaki positive-operator decomposition is applied
\cite{LiebYamazaki1958} to the driven Bose--Hubbard Hamiltonian.  Upon the
subtraction of a sum of positive operators, a quadratic bounding Hamiltonian
is left, and is subsequently reduced by a uniform displacement and a
Bogoliubov transformation~\cite{Bogoliubov1947,BlaizotRipka1986}.  An
admissible family of decomposition parameters is thereby delimited by the
resulting stability condition.  By the min--max principle, the quasiparticle
gap of the bounding Hamiltonian is then converted into a certified lower
threshold for the first excited energy of the full model.

From the construction there is obtained the rigorous lower bound of
Theorem~\ref{thm:lower-bound} in every dimension $d\geq1$ and its explicit
thermodynamic-limit energy density.  Proposition~\ref{prop:continuous-spectrum}
identifies the continuous excitation band of the bounding Hamiltonian, whilst
from the Bogoliubov vacuum there follow formulas that are exact within the Gaussian ground
state $|0_B\rangle$ of the bounding Hamiltonian $h_1$ for the order parameter,
filling, energy densities, particle-number fluctuations, anomalous averages,
and a drive-induced antibunching criterion.  Finally, the certified Temple
analysis of Sec.~\ref{sec:tightness} clarifies the finite-volume condition
under which this Gaussian ansatz can support a Temple correction. The
lower-bound property, \textit{per se}, does not depend upon that correction. Its
extensive energy excess makes that condition increasingly restrictive with
system size, independently of the quality of the lower energy-density bound.

\section{The Model and the Lieb--Yamazaki Procedure}

\subsection{Model Hamiltonian}

We consider an array of coupled single-mode microcavities containing nonlinear
media and driven coherently by external radiation.  In a frame rotating at the
driving frequency, the driven Bose--Hubbard Hamiltonian is
\begin{equation}
 H = \sum_{k=1}^{M} \left[
 \frac{V}{2}a_k^{\dagger 2}a_k^2
 -\hbar\Delta\omega_0\,a_k^\dagger a_k
 \right]
 -\frac{J}{2}\sum_{\langle k,l \rangle}
 \left(a_k^\dagger a_l+\text{h.c.}\right)
 +F\sum_{k=1}^{M}\left(a_k^\dagger+a_k\right).
 \label{eq:model-hamiltonian}
\end{equation}
Here, the bosonic operators obey
$[a_k,a_l^\dagger]=\delta_{kl}$.  The parameter $V$ is the on-site Kerr
nonlinearity, $J$ is the nearest-neighbor hopping amplitude, $F$ is the
coherent driving amplitude, and $\Delta\omega_0$ is the detuning between the
drive and the bare cavity mode.  The lattice contains $M$ sites and is endowed
with periodic boundary conditions.  The notation $\langle k,l\rangle$ denotes
nearest-neighbor pairs, counted once. Throughout we assume a repulsive
interaction $V>0$. A gauge transformation makes the hopping nonnegative,
so we take $J\geq0$; a global phase rotation likewise permits real $F$.

\subsection{Rescaling to Dimensionless Variables}

For $V>0$, we measure all energies in units of $V/2$ and introduce the
dimensionless quantities
\begin{equation}
 h=\frac{2H}{V},\qquad
 \nu_0=\frac{2\hbar\Delta\omega_0}{V},\qquad
 j=\frac{2J}{V},\qquad
 f=\frac{2F}{V},\qquad
 e=\frac{2E}{V}.
 \label{eq:dimensionless-parameters}
\end{equation}
The stationary Schr\"odinger equation then takes the form
\begin{equation}
 h\psi=e\psi.
 \label{eq:dimensionless-schrodinger}
\end{equation}
In these variables, the on-site interaction has unit coefficient, which makes
the algebraic structure used below explicit.

\subsection{Lieb--Yamazaki Decomposition}

Following the positive-operator strategy of Lieb and Yamazaki
\cite{LiebYamazaki1958}, we introduce at every lattice site
\begin{equation}
 B_k=a_k^\dagger a_k-K-g\left(a_k-a_k^\dagger\right),
 \label{eq:B-operator}
\end{equation}
where $K$ and $g$ are real variational parameters.  The operator
\begin{equation}
 w^2=\sum_{k=1}^{M}B_k^\dagger B_k
 \label{eq:w-square}
\end{equation}
is positive semidefinite. The remainder in the exact decomposition below is
\begin{equation}
 \begin{split}
 h_1={}&\sum_{k=1}^{M}\left[
 (2K-\nu_0-1)a_k^\dagger a_k
 +(f-g)(a_k^\dagger+a_k)
 +g^2(a_k-a_k^\dagger)^2
 \right] \\
 &-\frac{j}{2}\sum_{\langle k,l\rangle}
 \left(a_k^\dagger a_l+\text{h.c.}\right)-MK^2.
 \end{split}
 \label{eq:h1-before-shift}
\end{equation}

\begin{proposition}[Positive-Operator Decomposition]
\label{prop:ly-decomposition}
Let $B_k=a_k^\dagger a_k-K-g(a_k-a_k^\dagger)$ for real parameters $K$ and
$g$. Then the dimensionless Hamiltonian admits the exact decomposition
\begin{equation}
 h=w^2+h_1,
 \label{eq:ly-decomposition-prop}
\end{equation}
where $w^2=\sum_{k=1}^M B_k^\dagger B_k\geq0$ and $h_1$ is given explicitly
by Eq.~\eqref{eq:h1-before-shift}. Consequently, $h\geq h_1$ as an operator
inequality.
\end{proposition}

\begin{proof}[Proof of Proposition~\ref{prop:ly-decomposition}]
On the common invariant finite-particle core, direct expansion of
$B_k^\dagger B_k$, using $[a_k,a_l^\dagger]=\delta_{kl}$, gives
Eq.~\eqref{eq:ly-decomposition-prop} term by term. Moreover,
$\langle\psi,B_k^\dagger B_k\psi\rangle=\lVert B_k\psi\rVert^2\geq0$
for every $k$, so that summation over $k$ gives $w^2\geq0$; hence the asserted
inequality holds there as a quadratic-form inequality and extends to the
corresponding closed forms~\cite{ReedSimon1}. This is the positive-operator step
of the Lieb--Yamazaki construction~\cite{LiebYamazaki1958}.
\end{proof}

\subsection{Uniform c-number Shift}

The linear terms are eliminated by the real, spatially uniform displacement
$\lambda$,
\begin{equation}
 a_k=\widetilde{a}_k+\lambda,
 \qquad
 a_k^\dagger=\widetilde{a}_k^\dagger+\lambda.
 \label{eq:uniform-shift}
\end{equation}
While the original polaron application of Lieb and Yamazaki employs a
momentum-dependent coupling $g_\mu$, we adopt the uniform ansatz
$g=\text{const}$. This choice preserves the diagonal momentum-space structure
of the Bogoliubov transformation. It permits an analytic treatment of the
thermodynamic limit and closed-form expressions for the energy density and
excitation spectrum. The same framework also permits momentum-dependent
couplings.

With the dimensionless detuning $\nu_0$ of
Eq.~\eqref{eq:dimensionless-parameters}, define
\begin{equation}
 \widetilde{\nu}=2K-\nu_0-1.
 \label{eq:nu-tilde}
\end{equation}
Cancellation of the linear terms fixes the displacement to
\begin{equation}
 \lambda=-\frac{f-g}{\widetilde{\nu}-j},
 \qquad \widetilde{\nu}\ne j.
 \label{eq:optimal-shift}
\end{equation}
Substitution into Eq.~\eqref{eq:h1-before-shift} gives
\begin{equation}
 \begin{split}
 h_1={}&\sum_{k=1}^{M}\left[
 \widetilde{\nu}\,\widetilde{a}_k^\dagger\widetilde{a}_k
 +g^2(\widetilde{a}_k-\widetilde{a}_k^\dagger)^2
 \right] \\
 &-\frac{j}{2}\sum_{\langle k,l\rangle}
 \left(\widetilde{a}_k^\dagger\widetilde{a}_l+\text{h.c.}\right)
 -M\left(K^2+\frac{(f-g)^2}{\widetilde{\nu}-j}\right).
 \end{split}
 \label{eq:h1-after-shift}
\end{equation}
The last term is the complete c-number contribution, including the $-MK^2$
term generated by the Lieb--Yamazaki decomposition.

\subsection{Fourier Transformation}

Translation invariance permits the discrete Fourier representation
\begin{equation}
 \widetilde{a}_k=\frac{1}{\sqrt{M}}\sum_{\mu=0}^{M-1}
 b_\mu\exp\left(\frac{2\pi i k\mu}{M}\right),
 \label{eq:discrete-fourier}
\end{equation}
where $[b_\mu,b_{\mu'}^\dagger]=\delta_{\mu\mu'}$ and all momentum
indices are understood modulo $M$.  The quadratic lower-bound Hamiltonian
becomes
\begin{equation}
 \begin{split}
 h_1={}&\sum_{\mu=0}^{M-1}\left[
 \omega_\mu b_\mu^\dagger b_\mu
 +g^2\left(b_\mu b_{M-\mu}
 +b_\mu^\dagger b_{M-\mu}^\dagger\right)
 \right] \\
 &-M\left(K^2+\frac{(f-g)^2}{\widetilde{\nu}-j}+g^2\right),
 \end{split}
 \label{eq:h1-fourier}
\end{equation}
where
\begin{equation}
 \omega_\mu=\widetilde{\nu}
 -j\cos\left(\frac{2\pi\mu}{M}\right)-2g^2.
 \label{eq:omega-mu}
\end{equation}
A positive normal coefficient requires $\omega_\mu>0$ for every momentum
$\mu$.  The stronger stability condition needed by the Bogoliubov
transformation is obtained below.

\subsection{Bogoliubov Transformation}

We introduce new canonical bosonic modes $c_\mu$ through
\begin{equation}
 b_\mu=\frac{1}{\sqrt{1-A_\mu^2}}
 \left(c_\mu+A_\mu c_{M-\mu}^\dagger\right),
 \label{eq:bogoliubov-transformation}
\end{equation}
where $A_\mu=A_{M-\mu}$ is real and $|A_\mu|<1$.  Choosing $A_\mu$ to cancel
the anomalous terms diagonalizes Eq.~\eqref{eq:h1-fourier} and yields
\begin{equation}
 \begin{split}
 h_1={}&\sum_{\mu=0}^{M-1}\frac{1}{1-A_\mu^2}
 \left[\omega_\mu(1+A_\mu^2)+4g^2A_\mu\right]
 c_\mu^\dagger c_\mu \\
 &+g^2\sum_{\mu=0}^{M-1}A_\mu
 -M\left(K^2+\frac{(f-g)^2}{\widetilde{\nu}-j}+g^2\right).
 \end{split}
 \label{eq:h1-diagonal}
\end{equation}
The physical branch, continuously connected to $A_\mu=0$ as $g\to0$, is
\begin{equation}
 A_\mu=\frac{1}{2g^2}
 \left(-\omega_\mu+\sqrt{\omega_\mu^2-4g^4}\right).
 \label{eq:A-mu}
\end{equation}
\begin{lemma}[Stability Condition for Bogoliubov Transformation]
\label{lem:stability}
For $g\ne0$, the physical branch of the Bogoliubov transformation defined by
Eq.~\eqref{eq:bogoliubov-transformation} is canonical and diagonalizes a
positive quadratic mode Hamiltonian if and only if
\begin{equation}
 \omega_\mu>2g^2\quad\text{for all }\mu.
 \label{eq:stability-condition-lemma}
\end{equation}
For $j>0$, this condition is equivalent to the constraint on $K$ given by
Eq.~\eqref{eq:K-constraint-theorem}.
\end{lemma}

\begin{proof}[Proof of Lemma~\ref{lem:stability}]
The Bogoliubov transformation is canonical if and only if its coefficients
satisfy $|A_\mu|<1$ and the transformation preserves the commutation
relations. Reality of the square root alone permits either
$\omega_\mu>2g^2$ or $\omega_\mu<-2g^2$. The latter branch has a negative
coefficient of the number operator after diagonalization and is therefore
unbounded below. Positivity selects $\omega_\mu>2g^2$, for which the physical
root in Eq.~\eqref{eq:A-mu} also obeys $|A_\mu|<1$. For $j>0$, the minimum of $\omega_\mu$ occurs
at $\mu=0$, which yields the stated constraint on $K$. When $g=0$, one instead
takes $A_\mu=0$ directly; the positive-mode condition is $\omega_\mu>0$.
This is the standard bosonic Bogoliubov criterion
~\cite{Bogoliubov1947,BlaizotRipka1986}.
\end{proof}

\subsection{Extension to $d$ Dimensions}
\label{sec:d-dimensions}

The construction above does not depend on the dimension of the lattice. Let
$d\geq1$ and let $\Lambda=\{1,\dots,M\}^d$ be the periodic hypercubic lattice
of linear size $M$, with $|\Lambda|=M^d$ sites labeled by
$\mathbf{k}=(k_1,\dots,k_d)$. We consider
\begin{equation}
 H=\sum_{\mathbf{k}\in\Lambda}\left[
 \frac{V}{2}a_{\mathbf{k}}^{\dagger 2}a_{\mathbf{k}}^2
 -\hbar\Delta\omega_0\,a_{\mathbf{k}}^\dagger a_{\mathbf{k}}
 \right]
 -\frac{J}{2d}\sum_{\langle\mathbf{k},\mathbf{l}\rangle}
 \left(a_{\mathbf{k}}^\dagger a_{\mathbf{l}}+\text{h.c.}\right)
 +F\sum_{\mathbf{k}\in\Lambda}
 \left(a_{\mathbf{k}}^\dagger+a_{\mathbf{k}}\right),
 \label{eq:model-hamiltonian-d}
\end{equation}
which reduces to Eq.~\eqref{eq:model-hamiltonian} for $d=1$. The hopping
amplitude per spatial direction is thus $J/d$. With this normalization, the
hopping energy of the zero-momentum mode is independent of $d$. The
dimensionless quantities are defined as in
Eq.~\eqref{eq:dimensionless-parameters}.

The steps of the one-dimensional derivation carry over as follows.
\begin{enumerate}
\item The operator $B_{\mathbf{k}}$ of Eq.~\eqref{eq:B-operator} acts on a
single site, and the hopping term is not affected by the decomposition.
Proposition~\ref{prop:ly-decomposition} therefore holds verbatim, with sums
over $\Lambda$ and with the constant $-M^dK^2$ in $h_1$.
\item Every site belongs to $d$ nearest-neighbor pairs, each carrying the
coefficient $-j/(2d)$. The uniform displacement of
Eq.~\eqref{eq:uniform-shift} thus produces the same linear terms per site as
in one dimension, and Eq.~\eqref{eq:optimal-shift} for $\lambda$ is unchanged.
The c-number contribution becomes
$-M^d\left[K^2+(f-g)^2/(\widetilde{\nu}-j)\right]$.
\item The Fourier modes are labeled by
$\boldsymbol{\mu}=(\mu_1,\dots,\mu_d)\in\{0,\dots,M-1\}^d$,
\begin{equation}
 \widetilde{a}_{\mathbf{k}}=\frac{1}{M^{d/2}}\sum_{\boldsymbol{\mu}}
 b_{\boldsymbol{\mu}}
 \exp\left(\frac{2\pi i\,\mathbf{k}\cdot\boldsymbol{\mu}}{M}\right),
 \label{eq:discrete-fourier-d}
\end{equation}
and Eqs.~\eqref{eq:h1-fourier}--\eqref{eq:A-mu} hold with $\mu$ replaced by
$\boldsymbol{\mu}$, with $M-\mu$ replaced by $-\boldsymbol{\mu}$ (modulo $M$
in each component), with the constant $-M(\cdots)$ replaced by
$-M^d(\cdots)$, and with
\begin{equation}
 \omega_{\boldsymbol{\mu}}
 =\widetilde{\nu}+\epsilon_{\boldsymbol{\mu}}-2g^2,
 \qquad
 \epsilon_{\boldsymbol{\mu}}
 =-\frac{j}{d}\sum_{\alpha=1}^{d}\cos\left(\frac{2\pi\mu_\alpha}{M}\right).
 \label{eq:omega-d}
\end{equation}
Here $\epsilon_{\boldsymbol{\mu}}$ is the free-particle dispersion, with
$\epsilon_{\mathbf{0}}=-j$ for every $d$.
\item The excitation energies and the Bogoliubov coefficients are
\begin{equation}
 \Omega_{\boldsymbol{\mu}}
 =\sqrt{\omega_{\boldsymbol{\mu}}^2-4g^4},
 \qquad
 A_{\boldsymbol{\mu}}=\frac{1}{2g^2}
 \left(-\omega_{\boldsymbol{\mu}}+\Omega_{\boldsymbol{\mu}}\right),
 \label{eq:Omega-A-d}
\end{equation}
so that $g^2A_{\boldsymbol{\mu}}
=\tfrac12(\Omega_{\boldsymbol{\mu}}-\omega_{\boldsymbol{\mu}})$; for $g=0$
both sides vanish.
\item For $j\geq0$, the minimum of $\omega_{\boldsymbol{\mu}}$ is attained at
$\boldsymbol{\mu}=\mathbf{0}$, where all cosines equal one, and equals
$\widetilde{\nu}-j-2g^2$. The stability condition
$\omega_{\boldsymbol{\mu}}>2g^2$ of Lemma~\ref{lem:stability} is therefore
equivalent to the same constraint on $K$ in every dimension.
\end{enumerate}

\subsection{Ground-State Energy Lower Bound}

\begin{theorem}[Rigorous Lower Bound in $d$ Dimensions]
\label{thm:lower-bound}
Let $d\geq1$, let $H$ be the driven Bose--Hubbard Hamiltonian on
$\Lambda=\{1,\dots,M\}^d$ defined by Eq.~\eqref{eq:model-hamiltonian-d}, and
let $h=2H/V$ be the corresponding dimensionless Hamiltonian, with $j\geq0$.
For any real parameters $K$ and $g$ satisfying
\begin{equation}
 K>\frac{1}{2}\left(1+\nu_0+j+4g^2\right),
 \label{eq:K-constraint-theorem}
\end{equation}
the ground-state energy $E_0$ of $h$ satisfies the rigorous lower bound
\begin{equation}
 E_0\geq E^{(d)}_{0,l}=g^2\sum_{\boldsymbol{\mu}}A_{\boldsymbol{\mu}}
 -M^d\left(K^2+\frac{(f-g)^2}{\widetilde{\nu}-j}+g^2\right),
 \label{eq:lower-bound-theorem}
\end{equation}
where $\widetilde{\nu}=2K-\nu_0-1$, the sum runs over the $M^d$ modes
$\boldsymbol{\mu}\in\{0,\dots,M-1\}^d$, and $A_{\boldsymbol{\mu}}$ is given by
Eqs.~\eqref{eq:omega-d} and~\eqref{eq:Omega-A-d}. Equivalently,
\begin{equation}
 E^{(d)}_{0,l}=\frac12\sum_{\boldsymbol{\mu}}
 \left(\Omega_{\boldsymbol{\mu}}-\omega_{\boldsymbol{\mu}}\right)
 -M^d\left(K^2+\frac{(f-g)^2}{\widetilde{\nu}-j}+g^2\right).
 \label{eq:lower-bound-Omega}
\end{equation}
The constraint in Eq.~\eqref{eq:K-constraint-theorem} does not depend on $d$.
For $d=1$, Eq.~\eqref{eq:lower-bound-theorem} reads
$E^{(1)}_{0,l}=g^2\sum_{\mu=0}^{M-1}A_\mu
-M\left[K^2+(f-g)^2/(\widetilde{\nu}-j)+g^2\right]$, with $A_\mu$ given by
Eq.~\eqref{eq:A-mu}.
\end{theorem}

\begin{proof}[Proof of Theorem~\ref{thm:lower-bound}]
Proposition~\ref{prop:ly-decomposition} gives $h\geq h_1$ in the
quadratic-form sense. The variational (min--max) principle~\cite{ReedSimon4}
therefore implies $\inf\sigma(h)\geq\inf\sigma(h_1)$.
The Hamiltonian $h_1$ is quadratic in the bosonic operators and is
diagonalized by the Bogoliubov transformation in
Eq.~\eqref{eq:bogoliubov-transformation}, which is well-defined by
Lemma~\ref{lem:stability}. The vacuum of the $c_{\boldsymbol{\mu}}$ modes
minimizes the diagonalized $h_1$, and its lowest eigenvalue is given by
Eq.~\eqref{eq:lower-bound-theorem}; Eq.~\eqref{eq:lower-bound-Omega} follows
from Eq.~\eqref{eq:Omega-A-d}. The dimension enters this argument only
through the number of sites, $M^d$, and through the sum over modes: the
Lieb--Yamazaki decomposition is local, acting on one site at a time, and the
remaining steps were adapted to $\Lambda$ in Sec.~\ref{sec:d-dimensions}. The
proof is thus the same for every $d\geq1$.
\end{proof}

\subsection{Summary}

It follows that, in every dimension $d\geq1$, there is a variational family
of rigorous lower bounds for the ground-state energy, parameterized by $K$ and
$g$ subject to the stability constraint. In the following section, we take the
thermodynamic limit $M\to\infty$ to extract the energy density and excitation
spectrum.

\section{The Thermodynamic Limit}

\subsection{Strategy: From Finite Lattice to Thermodynamic Limit}

Theorem~\ref{thm:lower-bound} provides a rigorous lower bound
$E^{(d)}_{0,l}$ for every finite periodic lattice of $M^d$ sites, in any
dimension $d\geq1$. Its derivation uses the positive-operator inequality of
Proposition~\ref{prop:ly-decomposition} and the stability criterion of
Lemma~\ref{lem:stability}. We now prove the existence of the corresponding
energy density
\begin{equation}
 e^{(d)}_{0,l}=\lim_{M\to\infty}\frac{E^{(d)}_{0,l}}{M^d}.
 \label{eq:tdl-energy-density-definition}
\end{equation}
It is essential that passage to the thermodynamic limit preserve the
finite-volume lower-bound property. Let
$\mathbf{q}=(q_1,\dots,q_d)\in[-\pi,\pi]^d$ denote the momentum variable of
the $d$-dimensional Brillouin zone. If $F(\mathbf{q})$ is continuous and
$2\pi$-periodic in each component, the standard convergence theorem for
Riemann sums on the grids $2\pi\boldsymbol{\mu}/M$ gives
\begin{equation}
 \lim_{M\to\infty}\frac{1}{M^d}\sum_{\boldsymbol{\mu}}
 F\!\left(\frac{2\pi\boldsymbol{\mu}}{M}\right)
 =\int_{[-\pi,\pi]^d}\frac{d^dq}{(2\pi)^d}F(\mathbf{q}),
 \label{eq:A-riemann-limit}
\end{equation}
see, e.g., Ref.~\cite{Rudin1976}.

\subsection{Energy Density in the Thermodynamic Limit}

Dividing the finite-volume bound in
Eq.~\eqref{eq:lower-bound-theorem} by $M^d$ yields
\begin{equation}
 \frac{E^{(d)}_{0,l}}{M^d}
 =\frac{g^2}{M^d}\sum_{\boldsymbol{\mu}}A_{\boldsymbol{\mu}}
 -\left(K^2+\frac{(f-g)^2}{\widetilde{\nu}-j}+g^2\right).
 \label{eq:finite-volume-energy-density}
\end{equation}
Introduce the continuous functions
\begin{equation}
 \omega(\mathbf{q})=\widetilde{\nu}
 -\frac{j}{d}\sum_{\alpha=1}^{d}\cos q_\alpha-2g^2,
 \qquad
 \Omega(\mathbf{q})=\sqrt{\omega(\mathbf{q})^2-4g^4},
 \label{eq:omega-k-definition}
\end{equation}
and
\begin{equation}
 A(\mathbf{q})=\frac{1}{2g^2}
 \left(-\omega(\mathbf{q})+\Omega(\mathbf{q})\right).
 \label{eq:A-k}
\end{equation}
The function $A(\mathbf{q})$ is the root of
$g^2A^2+\omega(\mathbf{q})A+g^2=0$ with $|A|<1$, and
$g^2A(\mathbf{q})
=\tfrac12\left[\Omega(\mathbf{q})-\omega(\mathbf{q})\right]$.
For $g=0$, the expression $g^2A(\mathbf{q})$ is understood by its continuous
limit as $g\to0$, which is zero. By Eqs.~\eqref{eq:omega-d}
and~\eqref{eq:Omega-A-d},
$\omega_{\boldsymbol{\mu}}=\omega(2\pi\boldsymbol{\mu}/M)$,
$\Omega_{\boldsymbol{\mu}}=\Omega(2\pi\boldsymbol{\mu}/M)$, and, for
$g\ne0$, $A_{\boldsymbol{\mu}}=A(2\pi\boldsymbol{\mu}/M)$.

\begin{corollary}[Energy Density in the Thermodynamic Limit]
\label{cor:energy-density}
Let $d\geq1$ and $j\geq0$. For every admissible pair $(K,g)$ satisfying the
stability constraint in Eq.~\eqref{eq:K-constraint-theorem}, the
thermodynamic limit of the finite-volume lower-bound density exists and is
\begin{equation}
 e^{(d)}_{0,l}=\lim_{M\to\infty}\frac{E^{(d)}_{0,l}}{M^d}
 =g^2\int_{[-\pi,\pi]^d}\frac{d^dq}{(2\pi)^d}A(\mathbf{q})
 -\left(K^2+\frac{(f-g)^2}{\widetilde{\nu}-j}+g^2\right).
 \label{eq:energy-density}
\end{equation}
Equivalently,
\begin{equation}
 e^{(d)}_{0,l}
 =\frac12\int_{[-\pi,\pi]^d}\frac{d^dq}{(2\pi)^d}\,\Omega(\mathbf{q})
 -\left(K^2+\frac{(f-g)^2}{\widetilde{\nu}-j}
 +\frac{\widetilde{\nu}}{2}\right).
 \label{eq:energy-density-Omega}
\end{equation}
Without assuming existence of the exact thermodynamic limit, this expression
obeys
\begin{equation}
 \liminf_{M\to\infty}\frac{E_0(M)}{M^d}\geq e^{(d)}_{0,l}.
 \label{eq:exact-energy-liminf-bound}
\end{equation}
\end{corollary}

\begin{proof}[Proof of Corollary~\ref{cor:energy-density}]
For admissible parameters, $\omega(\mathbf{q})>2g^2$ for every $\mathbf{q}$,
and hence $\Omega(\mathbf{q})>0$ and $A(\mathbf{q})$ is continuous and
$2\pi$-periodic in each component. The Riemann-sum limit in
Eq.~\eqref{eq:A-riemann-limit} applies with $F=A$.
The remaining terms in Eq.~\eqref{eq:finite-volume-energy-density} are
independent of $M$, which proves Eq.~\eqref{eq:energy-density}.
Equation~\eqref{eq:energy-density-Omega} follows from
$g^2A=\tfrac12(\Omega-\omega)$, because the cosines integrate to zero and
the Brillouin-zone average of $\omega(\mathbf{q})$ is therefore
$\widetilde{\nu}-2g^2$. Moreover,
Theorem~\ref{thm:lower-bound} gives
$E_0(M)/M^d\geq E^{(d)}_{0,l}(M)/M^d$ for every $M$. Taking the lower limit
proves Eq.~\eqref{eq:exact-energy-liminf-bound}; if the exact energy density
exists, this reduces to the usual thermodynamic-limit inequality.
\end{proof}

\subsection{Continuous Bogoliubov Excitation Spectrum}

\begin{proposition}[Continuous Excitation Spectrum]
\label{prop:continuous-spectrum}
The diagonal coefficient in Eq.~\eqref{eq:h1-diagonal} is
$\Omega_{\boldsymbol{\mu}}=\sqrt{\omega_{\boldsymbol{\mu}}^2-4g^4}$. In the
thermodynamic limit these Bogoliubov excitation energies sample the
continuous dispersion
\begin{equation}
 \Omega(\mathbf{q})=\sqrt{\Bigl(\widetilde{\nu}
 -\frac{j}{d}\sum_{\alpha=1}^{d}\cos q_\alpha-2g^2\Bigr)^2-4g^4},
 \qquad \mathbf{q}\in[-\pi,\pi]^d.
 \label{eq:Omega-k}
\end{equation}
The stability condition $\omega_{\boldsymbol{\mu}}>2g^2$ becomes
$\omega(\mathbf{q})>2g^2$ for every $\mathbf{q}$, which for $j\geq0$ is
equivalent to the constraint on $K$ in
Eq.~\eqref{eq:K-constraint-theorem}, independently of $d$.
\end{proposition}

\begin{proof}[Proof of Proposition~\ref{prop:continuous-spectrum}]
Substitution of $A_{\boldsymbol{\mu}}$ into the coefficient of
$c_{\boldsymbol{\mu}}^\dagger c_{\boldsymbol{\mu}}$ in
Eq.~\eqref{eq:h1-diagonal} gives $\Omega_{\boldsymbol{\mu}}$ above, since
$A_{\boldsymbol{\mu}}$ solves
$g^2A_{\boldsymbol{\mu}}^2+\omega_{\boldsymbol{\mu}}A_{\boldsymbol{\mu}}
+g^2=0$. The momentum grids become dense in the Brillouin zone; continuity
of $\Omega(\mathbf{q})$ in the stable region then gives the claimed
continuum band. For $j\geq0$, the minimum of $\omega(\mathbf{q})$ occurs at
$\mathbf{q}=\mathbf{0}$ and equals $\widetilde{\nu}-j-2g^2$ in every
dimension, so uniform stability is precisely the constraint in
Eq.~\eqref{eq:K-constraint-theorem}.
\end{proof}

\subsection{Analysis of the Energy Density Integral}

For every admissible pair $(K,g)$, the strict inequality
$\omega(\mathbf{q})>2g^2$ ensures that the radicand in Eq.~\eqref{eq:A-k} is
positive throughout the compact Brillouin zone. Consequently,
$A(\mathbf{q})$ is smooth in $\mathbf{q}$, and the integral in
Eq.~\eqref{eq:energy-density} is well-defined. It may be evaluated
numerically for general parameters and analytically in special limits.

At the boundary of the admissible region, there may exist a critical momentum
$\mathbf{q}_c$ for which $\omega(\mathbf{q}_c)\to2g^2$. The square root
$\sqrt{\omega(\mathbf{q})^2-4g^4}$ then becomes nonanalytic. Such behavior
marks the loss of Bogoliubov stability of the bounding Hamiltonian; it is not
evidence of a phase transition of the full driven model. The rigorous bound
derived above applies in the open stable region; any extension to its boundary
requires a separate limiting argument.

\subsection{Summary}

We have thus rigorously established the thermodynamic limit of the
Lieb--Yamazaki lower bound in every dimension $d\geq1$. The lower-bound
density $e^{(d)}_{0,l}$ is given by a closed-form integral expression, and
the excitation spectrum is described by the continuous dispersion relation
$\Omega(\mathbf{q})$. In the following section, we derive macroscopic
observables from this framework.

\section{Excitation Spectrum and Observables}
\label{sec:spectrum}

\subsection{Order Parameter, Filling Fraction, and Energy Densities}
\label{sec:observables-part1}

Throughout this subsection, we take $d=1$; the extension to $d$ dimensions
is given in Remark~\ref{rem:observables-d}. The pair $(K,g)$ is admissible in
the sense of Eq.~\eqref{eq:K-constraint-theorem}, and $|0_B\rangle$ denotes the normalized
vacuum of the Bogoliubov modes of
Eq.~\eqref{eq:bogoliubov-transformation},
\begin{equation}
 c_\mu|0_B\rangle=0\quad\text{for all }\mu.
 \label{eq:bogoliubov-vacuum}
\end{equation}
By the proof of Theorem~\ref{thm:lower-bound}, $|0_B\rangle$ is the ground
state of the bounding Hamiltonian $h_1$; it depends on $K$ and $g$. All
expectation values below are taken in this state,
$\langle X\rangle=\langle0_B|X|0_B\rangle$, and all energies are measured in
the dimensionless units of Eq.~\eqref{eq:dimensionless-parameters}. We write
$\Omega_\mu=\sqrt{\omega_\mu^2-4g^4}$ as in
Proposition~\ref{prop:continuous-spectrum}.

\begin{lemma}[Two-Point Functions]
\label{lem:two-point}
In the state $|0_B\rangle$, $\langle b_\mu\rangle=0$ and
\begin{equation}
 \langle b_\mu^\dagger b_{\mu'}\rangle
 =\delta_{\mu\mu'}\frac{\omega_\mu-\Omega_\mu}{2\Omega_\mu},
 \qquad
 \langle b_\mu b_{\mu'}\rangle
 =-\delta_{\mu,M-\mu'}\frac{g^2}{\Omega_\mu},
 \label{eq:mode-two-point}
\end{equation}
with momentum indices understood modulo $M$. Consequently, the local moments
of the displaced operators are independent of the site $k$ and equal
\begin{equation}
 m\equiv\langle\widetilde{a}_k^\dagger\widetilde{a}_k\rangle
 =\frac1M\sum_{\mu=0}^{M-1}\frac{\omega_\mu-\Omega_\mu}{2\Omega_\mu},
 \qquad
 s\equiv\langle\widetilde{a}_k\widetilde{a}_k\rangle
 =-\frac{g^2}{M}\sum_{\mu=0}^{M-1}\frac{1}{\Omega_\mu},
 \label{eq:local-moments}
\end{equation}
while $\langle\widetilde{a}_k\rangle=0$.
\end{lemma}

\begin{proof}[Proof of Lemma~\ref{lem:two-point}]
Equation~\eqref{eq:bogoliubov-transformation} expresses $b_\mu$ as a linear
combination of $c_\mu$ and $c_{M-\mu}^\dagger$, so that
$\langle b_\mu\rangle=0$ by Eq.~\eqref{eq:bogoliubov-vacuum}. Using
$\langle c_\mu c_{\mu'}^\dagger\rangle=\delta_{\mu\mu'}$ and
$A_\mu=A_{M-\mu}$, we have
\begin{equation}
 \langle b_\mu^\dagger b_{\mu'}\rangle
 =\delta_{\mu\mu'}\frac{A_\mu^2}{1-A_\mu^2},
 \qquad
 \langle b_\mu b_{\mu'}\rangle
 =\delta_{\mu,M-\mu'}\frac{A_\mu}{1-A_\mu^2}.
 \label{eq:mode-two-point-A}
\end{equation}
Since $A_\mu$ in Eq.~\eqref{eq:A-mu} solves
$g^2A_\mu^2+\omega_\mu A_\mu+g^2=0$, it follows that
$1-A_\mu^2=-\Omega_\mu A_\mu/g^2$, which converts
Eq.~\eqref{eq:mode-two-point-A} into Eq.~\eqref{eq:mode-two-point}. For
$g=0$ one has $A_\mu=0$ and $\Omega_\mu=\omega_\mu$, and both sides of
Eq.~\eqref{eq:mode-two-point} vanish. Inserting
the Fourier representation in Eq.~\eqref{eq:discrete-fourier} yields
Eq.~\eqref{eq:local-moments}; the phases cancel in both sums, which proves
the independence of $k$.
\end{proof}

\begin{proposition}[Order Parameter]
\label{prop:order-parameter}
In the state $|0_B\rangle$, the expectation value of the annihilation operator
is the same at every site and equals the displacement,
\begin{equation}
 \langle a_k\rangle=\lambda=-\frac{f-g}{\widetilde{\nu}-j},
 \label{eq:order-parameter}
\end{equation}
where $\widetilde{\nu}=2K-\nu_0-1$ as in Eq.~\eqref{eq:nu-tilde}.
\end{proposition}

\begin{proof}[Proof of Proposition~\ref{prop:order-parameter}]
By Eq.~\eqref{eq:uniform-shift},
$\langle a_k\rangle=\langle\widetilde{a}_k\rangle+\lambda$, and
$\langle\widetilde{a}_k\rangle=0$ by Lemma~\ref{lem:two-point}. The value of
$\lambda$ is given by Eq.~\eqref{eq:optimal-shift}.
\end{proof}

\begin{proposition}[Filling Fraction]
\label{prop:filling-fraction}
In the state $|0_B\rangle$, the mean number of particles per site is
\begin{equation}
 \rho\equiv\langle a_k^\dagger a_k\rangle
 =\lambda^2+\frac1M\sum_{\mu=0}^{M-1}
 \frac{\omega_\mu-\Omega_\mu}{2\Omega_\mu},
 \label{eq:filling-fraction}
\end{equation}
with $\lambda$ given by Eq.~\eqref{eq:order-parameter}.
\end{proposition}

\begin{proof}[Proof of Proposition~\ref{prop:filling-fraction}]
Since $\lambda$ is real, Eq.~\eqref{eq:uniform-shift} gives
$a_k^\dagger a_k=\lambda^2
+\lambda(\widetilde{a}_k+\widetilde{a}_k^\dagger)
+\widetilde{a}_k^\dagger\widetilde{a}_k$. The linear terms have vanishing
expectation, and the last term equals $m$ by Lemma~\ref{lem:two-point}.
\end{proof}

The first term in Eq.~\eqref{eq:filling-fraction} is the coherent
contribution of the displacement, while the second is the depletion generated
by the anomalous terms of $h_1$; the latter vanishes for $g=0$.

\begin{proposition}[Kinetic Energy Density]
\label{prop:kinetic-energy-density}
Let
\begin{equation}
 T=-\frac{j}{2}\sum_{\langle k,l\rangle}
 \left(a_k^\dagger a_l+\text{h.c.}\right)
 \label{eq:kinetic-operator}
\end{equation}
be the hopping part of the dimensionless Hamiltonian $h$. In the state
$|0_B\rangle$,
\begin{equation}
 \frac{\langle T\rangle}{M}
 =\epsilon_0\lambda^2+\frac1M\sum_{\mu=0}^{M-1}
 \epsilon_\mu\frac{\omega_\mu-\Omega_\mu}{2\Omega_\mu},
 \qquad
 \epsilon_\mu=-j\cos\left(\frac{2\pi\mu}{M}\right),
 \label{eq:kinetic-energy-density}
\end{equation}
where $\epsilon_\mu$ is the free-particle dispersion and $\epsilon_0=-j$.
\end{proposition}

\begin{proof}[Proof of Proposition~\ref{prop:kinetic-energy-density}]
Let $\alpha_\mu$ denote the Fourier modes of the undisplaced operators $a_k$,
defined as in Eq.~\eqref{eq:discrete-fourier}. Then
$T=\sum_\mu\epsilon_\mu\alpha_\mu^\dagger\alpha_\mu$, and
Eq.~\eqref{eq:uniform-shift} gives
$\alpha_\mu=b_\mu+\lambda\sqrt{M}\,\delta_{\mu,0}$. By
Lemma~\ref{lem:two-point},
$\langle\alpha_\mu^\dagger\alpha_\mu\rangle
=\lambda^2M\delta_{\mu,0}+(\omega_\mu-\Omega_\mu)/(2\Omega_\mu)$.
Summation over $\mu$ and division by $M$ prove the claim.
\end{proof}

\begin{proposition}[On-Site Energy Density]
\label{prop:potential-energy-density}
Let
\begin{equation}
 V_k=a_k^{\dagger2}a_k^2-\nu_0a_k^\dagger a_k+f\left(a_k^\dagger+a_k\right)
 \label{eq:potential-operator}
\end{equation}
be the on-site part of $h$, so that $h=T+\sum_{k=1}^MV_k$. In the state
$|0_B\rangle$,
\begin{equation}
 \langle V_k\rangle
 =\lambda^4+2\lambda^2(2m+s)+2m^2+s^2
 -\nu_0\left(\lambda^2+m\right)+2f\lambda,
 \label{eq:potential-energy-density}
\end{equation}
where $m$ and $s$ are the local moments in Eq.~\eqref{eq:local-moments}.
This expression is exact within the Gaussian ground state $|0_B\rangle$ of
the bounding Hamiltonian $h_1$.
\end{proposition}

\begin{proof}[Proof of Proposition~\ref{prop:potential-energy-density}]
The detuning and driving terms follow from
Propositions~\ref{prop:order-parameter} and~\ref{prop:filling-fraction}. For
the interaction term, Eq.~\eqref{eq:uniform-shift} gives
\begin{equation}
 a_k^{\dagger2}a_k^2
 =\left(\widetilde{a}_k^{\dagger2}+2\lambda\widetilde{a}_k^\dagger
 +\lambda^2\right)
 \left(\widetilde{a}_k^2+2\lambda\widetilde{a}_k+\lambda^2\right).
 \label{eq:quartic-expansion}
\end{equation}
The state $|0_B\rangle$ is the vacuum of operators that are linear in
$\widetilde{a}_k$ and $\widetilde{a}_k^\dagger$, hence it is quasi-free and
Wick's theorem applies~\cite{BlaizotRipka1986}. Expectation values of odd
monomials vanish, and
$\langle\widetilde{a}_k^{\dagger2}\widetilde{a}_k^2\rangle=2m^2+s^2$ because
$s$ is real. The remaining even terms of Eq.~\eqref{eq:quartic-expansion}
contribute $\lambda^4+4\lambda^2m+2\lambda^2s$.
\end{proof}

\begin{remark}[Strong-Condensate Limit]
\label{rem:strong-condensate}
If the displacement dominates the fluctuations,
$\lambda^2\gg\max(m,|s|)$, the terms $2m^2+s^2$ in
Eq.~\eqref{eq:potential-energy-density} may be neglected. To first order in
the fluctuation moments one then has the approximation
\begin{equation}
 \langle V_k\rangle\approx
 \lambda^4-\nu_0\lambda^2+2f\lambda
 +\left(4\lambda^2-\nu_0\right)m+2\lambda^2s.
 \label{eq:potential-strong-condensate}
\end{equation}
In contrast with Eq.~\eqref{eq:potential-energy-density}, this is not an
identity; it is valid only in the strong-condensate limit.
\end{remark}

\begin{corollary}[Observables in the Thermodynamic Limit]
\label{cor:observables-tdl}
For every admissible pair $(K,g)$, the limits $M\to\infty$ of the quantities
above exist. With $\omega(q)$ and $\Omega(q)$ as in
Eq.~\eqref{eq:omega-k-definition} for $d=1$,
\begin{align}
 m_\infty&=\int_{-\pi}^{\pi}\frac{dq}{2\pi}
 \frac{\omega(q)-\Omega(q)}{2\Omega(q)},
 &
 s_\infty&=-g^2\int_{-\pi}^{\pi}\frac{dq}{2\pi}\frac{1}{\Omega(q)},
 \label{eq:local-moments-tdl}
 \\
 \rho_\infty&=\lambda^2+m_\infty,
 &
 \lim_{M\to\infty}\frac{\langle T\rangle}{M}
 &=-j\lambda^2-j\int_{-\pi}^{\pi}\frac{dq}{2\pi}
 \cos q\,\frac{\omega(q)-\Omega(q)}{2\Omega(q)},
 \label{eq:observables-tdl}
\end{align}
and the limit of $\langle V_k\rangle$ is given by
Eq.~\eqref{eq:potential-energy-density} with $m$ and $s$ replaced by
$m_\infty$ and $s_\infty$. The order parameter $\lambda$ is independent of
$M$.
\end{corollary}

\begin{proof}[Proof of Corollary~\ref{cor:observables-tdl}]
In the stable region, $\Omega(q)>0$ for every $q$, so the integrands are
continuous and $2\pi$-periodic. The finite-volume expressions are their
Riemann sums on the grid $q=2\pi\mu/M$, and the convergence argument used for
Eq.~\eqref{eq:A-riemann-limit} applies~\cite{Rudin1976}.
\end{proof}

\begin{remark}[Extension to $d$ Dimensions]
\label{rem:observables-d}
On the lattice $\Lambda=\{1,\dots,M\}^d$ of Sec.~\ref{sec:d-dimensions}, the
results of this subsection hold with the following replacements:
\begin{enumerate}
\item the mode averages $\frac1M\sum_{\mu=0}^{M-1}$ become
$\frac{1}{M^d}\sum_{\boldsymbol{\mu}}$, with $\omega_\mu$ and $\Omega_\mu$
replaced by $\omega_{\boldsymbol{\mu}}$ and $\Omega_{\boldsymbol{\mu}}$ of
Eqs.~\eqref{eq:omega-d} and~\eqref{eq:Omega-A-d}, and with
$\delta_{\mu,M-\mu'}$ in Eq.~\eqref{eq:mode-two-point} replaced by
$\delta_{\boldsymbol{\mu},-\boldsymbol{\mu}'}$;
\item the free-particle dispersion
$\epsilon_\mu=-j\cos(2\pi\mu/M)$ becomes $\epsilon_{\boldsymbol{\mu}}$ of
Eq.~\eqref{eq:omega-d}, the hopping operator in
Eq.~\eqref{eq:kinetic-operator} carries the coefficient $-j/(2d)$, and
$\langle T\rangle/M$ becomes $\langle T\rangle/M^d$;
\item the Brillouin-zone integrals and the cosine in
Eq.~\eqref{eq:observables-tdl} become
\begin{equation}
 \int_{-\pi}^{\pi}\frac{dq}{2\pi}\;\longrightarrow\;
 \int_{[-\pi,\pi]^d}\frac{d^dq}{(2\pi)^d},
 \qquad
 \cos q\;\longrightarrow\;\frac1d\sum_{\alpha=1}^{d}\cos q_\alpha.
 \label{eq:observables-d-rules}
\end{equation}
\end{enumerate}
The order parameter $\lambda$ and the expressions of
Propositions~\ref{prop:filling-fraction}
and~\ref{prop:potential-energy-density} and of
Remark~\ref{rem:strong-condensate} in terms of $\lambda$, $m$, and $s$ do not
change, because they are local and $\lambda$ does not depend on $d$. All
integrands remain continuous, since $\Omega(\mathbf{q})>0$ in the stable
region.
\end{remark}

\begin{remark}[Variational Upper Bound]
\label{rem:upper-bound}
Since $h=T+\sum_kV_k$ and $|0_B\rangle$ is a normalized trial state,
Propositions~\ref{prop:kinetic-energy-density}
and~\ref{prop:potential-energy-density} also give a variational upper bound
on the ground-state energy. Combined with Theorem~\ref{thm:lower-bound}
for $d=1$,
\begin{equation}
 \frac{E^{(1)}_{0,l}}{M}\leq\frac{E_0}{M}\leq
 \frac{\langle T\rangle}{M}+\langle V_k\rangle,
 \label{eq:two-sided-bound}
\end{equation}
and by Proposition~\ref{prop:ly-decomposition} the difference between the
two outer expressions is $\langle B_k^\dagger B_k\rangle\geq0$. In $d$
dimensions the same statement holds with $E^{(d)}_{0,l}$ and $M^d$ in place
of $E^{(1)}_{0,l}$ and $M$.
\end{remark}

\subsection{Particle-Number Variance and Correlations}
\label{sec:observables-part2}

We now turn from one-body expectation values to local two-body quantities in
the state $|0_B\rangle$: the variance of the on-site particle number, the
second-order correlation function, and the anomalous averages. Control of
these fluctuations is a prerequisite for assessing the quality of the lower
bound in Sec.~\ref{sec:tightness}. As in Sec.~\ref{sec:observables-part1}, we
take $d=1$; since every result below is expressed through $\lambda$, $m$, and
$s$, Remark~\ref{rem:observables-d} applies without change. Because
$|0_B\rangle$ is quasi-free, all formulas of this subsection follow from the
two-point functions of Lemma~\ref{lem:two-point} by Wick's theorem and are
exact, with the exception of Remark~\ref{rem:fluctuations-strong-condensate}.

We write $n_k=a_k^\dagger a_k$ and recall from the proof of
Proposition~\ref{prop:potential-energy-density} that
\begin{equation}
 \langle a_k^{\dagger2}a_k^2\rangle
 =\lambda^4+2\lambda^2(2m+s)+2m^2+s^2.
 \label{eq:quartic-moment}
\end{equation}

\begin{lemma}[Sign of the Local Moments]
\label{lem:moment-signs}
For $g\ne0$, the local moments of Eq.~\eqref{eq:local-moments} satisfy
\begin{equation}
 m>0,\qquad s<0,\qquad m+s<0.
 \label{eq:moment-signs}
\end{equation}
For $g=0$, $m=s=0$.
\end{lemma}

\begin{proof}[Proof of Lemma~\ref{lem:moment-signs}]
For $g\ne0$ and admissible parameters, $\omega_\mu>2g^2>0$ and
$0<\Omega_\mu<\omega_\mu$, which gives $m>0$ and $s<0$ term by term.
Moreover,
\begin{equation}
 m+s=\frac1M\sum_{\mu=0}^{M-1}
 \frac{\omega_\mu-2g^2-\Omega_\mu}{2\Omega_\mu},
 \label{eq:m-plus-s}
\end{equation}
and
$\Omega_\mu=\sqrt{(\omega_\mu-2g^2)(\omega_\mu+2g^2)}>\omega_\mu-2g^2$, so
every term of the sum is negative. For $g=0$, $\Omega_\mu=\omega_\mu$ and
both moments vanish.
\end{proof}

\begin{proposition}[Particle-Number Variance]
\label{prop:number-variance}
In the state $|0_B\rangle$, the variance of the on-site particle number is
\begin{equation}
 \operatorname{Var}(n_k)\equiv\langle n_k^2\rangle-\langle n_k\rangle^2
 =\rho+2\lambda^2(m+s)+m^2+s^2,
 \label{eq:number-variance}
\end{equation}
where $\rho=\lambda^2+m$ is the filling fraction of
Eq.~\eqref{eq:filling-fraction}. This expression is exact within
$|0_B\rangle$, the Gaussian ground state of $h_1$.
\end{proposition}

\begin{proof}[Proof of Proposition~\ref{prop:number-variance}]
The commutation relation $[a_k,a_k^\dagger]=1$ gives the operator identity
$n_k^2=a_k^{\dagger2}a_k^2+n_k$, hence
$\langle n_k^2\rangle=\langle a_k^{\dagger2}a_k^2\rangle+\rho$. Subtracting
$\rho^2=\lambda^4+2\lambda^2m+m^2$ from Eq.~\eqref{eq:quartic-moment} yields
Eq.~\eqref{eq:number-variance}.
\end{proof}

Two limiting cases illustrate Eq.~\eqref{eq:number-variance}. For $g=0$,
$m=s=0$ and $\operatorname{Var}(n_k)=\lambda^2=\rho$, the Poissonian value of
a coherent state. For $\lambda=0$,
$\operatorname{Var}(n_k)=m(1+m)+s^2>\rho$, the super-Poissonian value of a
squeezed vacuum. In general, Lemma~\ref{lem:moment-signs} shows that the
term $2\lambda^2(m+s)$ is negative for $g\ne0$, so that the number
fluctuations are sub-Poissonian, $\operatorname{Var}(n_k)<\rho$, if and only
if
\begin{equation}
 \lambda^2>\frac{m^2+s^2}{2|m+s|}.
 \label{eq:sub-poissonian-condition}
\end{equation}

\begin{proposition}[Second-Order Correlation Function]
\label{prop:g2}
In the state $|0_B\rangle$ with $\rho>0$, the local second-order correlation
function is
\begin{equation}
 g^{(2)}(0)\equiv
 \frac{\langle a_k^{\dagger2}a_k^2\rangle}{\langle a_k^\dagger a_k\rangle^2}
 =\frac{\lambda^4+2\lambda^2(2m+s)+2m^2+s^2}{\left(\lambda^2+m\right)^2}
 =1+\frac{2\lambda^2(m+s)+m^2+s^2}{\left(\lambda^2+m\right)^2}.
 \label{eq:g2}
\end{equation}
This expression is exact within $|0_B\rangle$, the Gaussian ground state of
$h_1$, and
$g^{(2)}(0)-1=[\operatorname{Var}(n_k)-\rho]/\rho^2$.
\end{proposition}

\begin{proof}[Proof of Proposition~\ref{prop:g2}]
The first equality is Eq.~\eqref{eq:quartic-moment} divided by $\rho^2$, with
$\rho$ given by Proposition~\ref{prop:filling-fraction}; the second follows
by subtracting $\rho^2$ in the numerator. The last statement is a
consequence of Proposition~\ref{prop:number-variance}.
\end{proof}

Values $g^{(2)}(0)>1$ correspond to bunching, $g^{(2)}(0)=1$ to Poissonian
statistics, and $g^{(2)}(0)<1$ to antibunching. For $g=0$ and
$\lambda\ne0$, Eq.~\eqref{eq:g2} gives $g^{(2)}(0)=1$, as for a coherent
state. For $\lambda=0$ and $g\ne0$, it gives $g^{(2)}(0)=2+s^2/m^2$, which
exceeds $3$ because $|s|>m$ by Lemma~\ref{lem:moment-signs}; this is the
strong bunching of a squeezed vacuum. Since $m+s<0$ for $g\ne0$, a
sufficiently large displacement reverses the sign of $g^{(2)}(0)-1$: the
state $|0_B\rangle$ is antibunched precisely when
Eq.~\eqref{eq:sub-poissonian-condition} holds. As $\lambda$ is proportional
to $f-g$ by Eq.~\eqref{eq:order-parameter}, this antibunching is induced by
the coherent drive.  The use of $g^{(2)}(0)$ as a blockade and nonclassicality
diagnostic parallels recent studies of dynamical, multi-emitter, and
squeezing-induced bosonic blockade~\cite{GhoshLiew2019,Trivedi2019,
KowalewskaKudlaszyk2019}.

\begin{proposition}[Anomalous Averages]
\label{prop:anomalous-averages}
In the state $|0_B\rangle$,
\begin{equation}
 \langle a_k^2\rangle=\langle a_k^{\dagger2}\rangle=\lambda^2+s.
 \label{eq:anomalous-averages}
\end{equation}
\end{proposition}

\begin{proof}[Proof of Proposition~\ref{prop:anomalous-averages}]
By Eq.~\eqref{eq:uniform-shift},
$a_k^2=\widetilde{a}_k^2+2\lambda\widetilde{a}_k+\lambda^2$. The linear term
has vanishing expectation and
$\langle\widetilde{a}_k^2\rangle=s$ by Lemma~\ref{lem:two-point}. The two
averages coincide because $\lambda$ and $s$ are real.
\end{proof}

The difference $\langle a_k^2\rangle-\langle a_k\rangle^2=s$ is the anomalous
average of the fluctuations. It originates from the pairing of the modes
$\mu$ and $M-\mu$ in Eq.~\eqref{eq:mode-two-point}; it is nonzero for
$g\ne0$ and vanishes for $g=0$.

\begin{remark}[Strong-Condensate Limit of the Fluctuations]
\label{rem:fluctuations-strong-condensate}
If $\lambda^2\gg\max(m,|s|)$, the terms $m^2+s^2$ in
Eqs.~\eqref{eq:number-variance} and~\eqref{eq:g2} may be neglected. To first
order in the fluctuation moments one then has the approximations
\begin{equation}
 \operatorname{Var}(n_k)\approx\lambda^2\left(1+2m+2s\right)+m,
 \qquad
 g^{(2)}(0)\approx1+\frac{2(m+s)}{\lambda^2}
 +\mathcal{O}\!\left(\lambda^{-4}\right).
 \label{eq:fluctuations-strong-condensate}
\end{equation}
In contrast with Propositions~\ref{prop:number-variance}
and~\ref{prop:g2}, these are not identities; they are valid only in the
strong-condensate limit. By Lemma~\ref{lem:moment-signs}, in this limit
$g^{(2)}(0)$ approaches unity from below whenever $g\ne0$.
\end{remark}

\begin{corollary}[Fluctuations in the Thermodynamic Limit]
\label{cor:fluctuations-tdl}
For every admissible pair $(K,g)$, the limits $M\to\infty$ of
$\operatorname{Var}(n_k)$, $g^{(2)}(0)$ (for $\rho_\infty>0$), and
$\langle a_k^2\rangle$ exist. They are given by
Eqs.~\eqref{eq:number-variance}, \eqref{eq:g2},
and~\eqref{eq:anomalous-averages} with $m$ and $s$ replaced by the integrals
$m_\infty$ and $s_\infty$ of Eq.~\eqref{eq:local-moments-tdl}. The
inequalities of Lemma~\ref{lem:moment-signs} hold for $m_\infty$ and
$s_\infty$ as well.
\end{corollary}

\begin{proof}[Proof of Corollary~\ref{cor:fluctuations-tdl}]
The three quantities are continuous functions of $(m,s)$ at fixed $\lambda$,
and $m\to m_\infty$, $s\to s_\infty$ by
Corollary~\ref{cor:observables-tdl}. The integrands of $m_\infty$,
$s_\infty$, and $m_\infty+s_\infty$ have the signs established pointwise in
the proof of Lemma~\ref{lem:moment-signs}.
\end{proof}

\subsection{Boundary Slope of the Bounding Hamiltonian}
\label{sec:sound-velocity}

We finally examine the long-wavelength behavior of the continuous band
$\Omega(\mathbf{q})$ of Proposition~\ref{prop:continuous-spectrum}, with
$\omega(\mathbf{q})$ and $\Omega(\mathbf{q})$ as in
Eqs.~\eqref{eq:omega-k-definition} and~\eqref{eq:Omega-k}. In contrast with
Secs.~\ref{sec:observables-part1} and~\ref{sec:observables-part2}, this
subsection is formulated directly for arbitrary $d\geq1$. We assume $j>0$;
for $j=0$ the band is flat. Because $\Omega(\mathbf{q})$ is a function of the
continuous momentum $\mathbf{q}$, every statement below already refers to the
thermodynamic limit, and no separate limiting corollary is needed. We use
\begin{equation}
 \omega(\mathbf{0})=\widetilde{\nu}-j-2g^2,
 \qquad
 \Omega(\mathbf{0})=\sqrt{\omega(\mathbf{0})^2-4g^4}
 =\sqrt{\left(\widetilde{\nu}-j\right)
 \left(\widetilde{\nu}-j-4g^2\right)}.
 \label{eq:omega-Omega-zero}
\end{equation}

\begin{proposition}[Boundary Slope]
\label{prop:sound-velocity}
Let $d\geq1$ and $j>0$.
\begin{enumerate}
\item In the stable region $\widetilde{\nu}-j>4g^2$, the excitation spectrum
has a gap, attained at $\mathbf{q}=\mathbf{0}$:
\begin{equation}
 \min_{\mathbf{q}}\Omega(\mathbf{q})=\Omega(\mathbf{0})
 =\sqrt{\left(\widetilde{\nu}-j-2g^2\right)^2-4g^4}>0.
 \label{eq:gap}
\end{equation}
As $\mathbf{q}\to\mathbf{0}$,
\begin{equation}
 \Omega(\mathbf{q})=\Omega(\mathbf{0})+\frac{|\mathbf{q}|^2}{2m^*}
 +\mathcal{O}\!\left(|\mathbf{q}|^4\right),
 \qquad
 \frac{1}{m^*}
 =\frac{\left(\widetilde{\nu}-j-2g^2\right)j}{d\,\Omega(\mathbf{0})},
 \label{eq:effective-mass}
\end{equation}
so that the dispersion is quadratic and the ratio
$\Omega(\mathbf{q})/|\mathbf{q}|$ diverges as $\mathbf{q}\to\mathbf{0}$:
there is no finite linear boundary slope.
\item On the boundary of the stable region, $\widetilde{\nu}-j=4g^2$ with
$g\ne0$, the function $\Omega(\mathbf{q})$ of Eq.~\eqref{eq:Omega-k} remains
well defined and
\begin{equation}
 \Omega(\mathbf{q})^2=\frac{j}{d}\,S(\mathbf{q})
 \left(4g^2+\frac{j}{d}\,S(\mathbf{q})\right),
 \qquad
 S(\mathbf{q})=\sum_{\alpha=1}^{d}\left(1-\cos q_\alpha\right).
 \label{eq:boundary-dispersion}
\end{equation}
The gap is closed and the dispersion is linear at small momenta,
\begin{equation}
 \Omega(\mathbf{q})=c\,|\mathbf{q}|
 +\mathcal{O}\!\left(|\mathbf{q}|^3\right),
 \qquad
 c=|g|\sqrt{\frac{2j}{d}}.
 \label{eq:sound-velocity}
\end{equation}
\end{enumerate}
\end{proposition}

\begin{proof}[Proof of Proposition~\ref{prop:sound-velocity}]
By Eq.~\eqref{eq:omega-k-definition},
$\omega(\mathbf{q})=\omega(\mathbf{0})+\frac{j}{d}S(\mathbf{q})$ with
$S(\mathbf{q})\geq0$, and
$S(\mathbf{q})=\tfrac12|\mathbf{q}|^2+\mathcal{O}(|\mathbf{q}|^4)$. Hence
\begin{equation}
 \Omega(\mathbf{q})^2=\Omega(\mathbf{0})^2
 +\frac{2\omega(\mathbf{0})j}{d}\,S(\mathbf{q})
 +\frac{j^2}{d^2}\,S(\mathbf{q})^2
 =\Omega(\mathbf{0})^2+\frac{\omega(\mathbf{0})j}{d}\,|\mathbf{q}|^2
 +\mathcal{O}\!\left(|\mathbf{q}|^4\right).
 \label{eq:Omega-squared-expansion}
\end{equation}
In the stable region $\omega(\mathbf{0})>2g^2\geq0$, so the right-hand side
of the first equality is increasing in $S$ and is minimal at
$\mathbf{q}=\mathbf{0}$, which gives Eq.~\eqref{eq:gap}. Since
$\Omega(\mathbf{0})>0$, the square root is smooth near
$\mathbf{q}=\mathbf{0}$, and its expansion gives
Eq.~\eqref{eq:effective-mass}. On the boundary,
$\omega(\mathbf{0})=2g^2$ and $\Omega(\mathbf{0})=0$, and the first equality
in Eq.~\eqref{eq:Omega-squared-expansion} reduces to
Eq.~\eqref{eq:boundary-dispersion}. Then
$\Omega(\mathbf{q})^2=\frac{2g^2j}{d}|\mathbf{q}|^2
+\mathcal{O}(|\mathbf{q}|^4)$, and taking the square root gives
Eq.~\eqref{eq:sound-velocity}.
\end{proof}

The two regimes are connected by
Eq.~\eqref{eq:Omega-squared-expansion}, which has the relativistic form
$\Omega(\mathbf{q})^2\approx\Omega(\mathbf{0})^2+v^2|\mathbf{q}|^2$ with
$v^2=\omega(\mathbf{0})j/d$, so that $m^*v^2=\Omega(\mathbf{0})$. In the
stable region the excitations of the bounding Hamiltonian are massive
quasiparticles with effective mass $m^*$; the dispersion is quadratic for
$|\mathbf{q}|\ll\Omega(\mathbf{0})/v$ and approximately linear with slope $v$
above this scale. As the boundary is approached, $m^*\to0$, $v\to c$, and the
quadratic window shrinks to the point $\mathbf{q}=\mathbf{0}$; the limits
$\mathbf{q}\to\mathbf{0}$ and $\widetilde{\nu}-j\to4g^2$ therefore do not
commute. The velocity $c=|g|\sqrt{2j/d}$ depends only on the hopping
amplitude $j$, the parameter $g$, and the dimension; it does not depend on
the detuning $\nu_0$ or on the drive amplitude $f$, because on the boundary
$K$ is fixed by $\widetilde{\nu}=j+4g^2$.

Two limitations should be kept in mind. First, $K$ and $g$ are parameters of
the Lieb--Yamazaki decomposition and not of the model, so the gap, $m^*$, and
$c$ characterize the bounding Hamiltonian $h_1$ for a given choice of
$(K,g)$; the boundary $\widetilde{\nu}-j=4g^2$ is the edge of the admissible
region and not a phase boundary of the driven Bose--Hubbard model. Second,
Theorem~\ref{thm:lower-bound} and the results of
Secs.~\ref{sec:observables-part1} and~\ref{sec:observables-part2} are proved
in the open stable region only. On the boundary the integrands of
Eq.~\eqref{eq:local-moments-tdl} behave as $1/(c|\mathbf{q}|)$ near
$\mathbf{q}=\mathbf{0}$, which is integrable for $d\geq2$ but not for $d=1$,
where $m_\infty$ and $|s_\infty|$ diverge logarithmically as the boundary is
approached.

\begin{remark}[Comparison with the Case $g=0$]
\label{rem:sound-velocity-g0}
For $g=0$ the spectrum is $\Omega(\mathbf{q})=\omega(\mathbf{q})$, the stable
region is $\widetilde{\nu}-j>0$, and Eq.~\eqref{eq:effective-mass} gives the
free lattice mass $1/m^*=j/d$. On the boundary $\widetilde{\nu}=j$ one has
$\Omega(\mathbf{q})=\frac{j}{d}S(\mathbf{q})
=\frac{j}{2d}|\mathbf{q}|^2+\mathcal{O}(|\mathbf{q}|^4)$: the gap closes,
but the dispersion remains quadratic and the boundary slope vanishes, in
agreement with Eq.~\eqref{eq:sound-velocity}. Moreover, the displacement
$\lambda$ of Eq.~\eqref{eq:optimal-shift} diverges there unless $f=0$. For
$g\ne0$, in contrast, $\lambda=-(f-g)/(4g^2)$ is finite on the boundary and
the gapless excitations have the linear dispersion of
Eq.~\eqref{eq:sound-velocity}. A finite linear boundary slope at the edge of
the stable region is thus a qualitative effect of the pairing terms generated
by $g\ne0$.
\end{remark}

\subsection{Numerical Illustration of the Spectrum and Local Correlations}

\begin{figure}[p]
 \centering
 \includegraphics[width=\columnwidth]{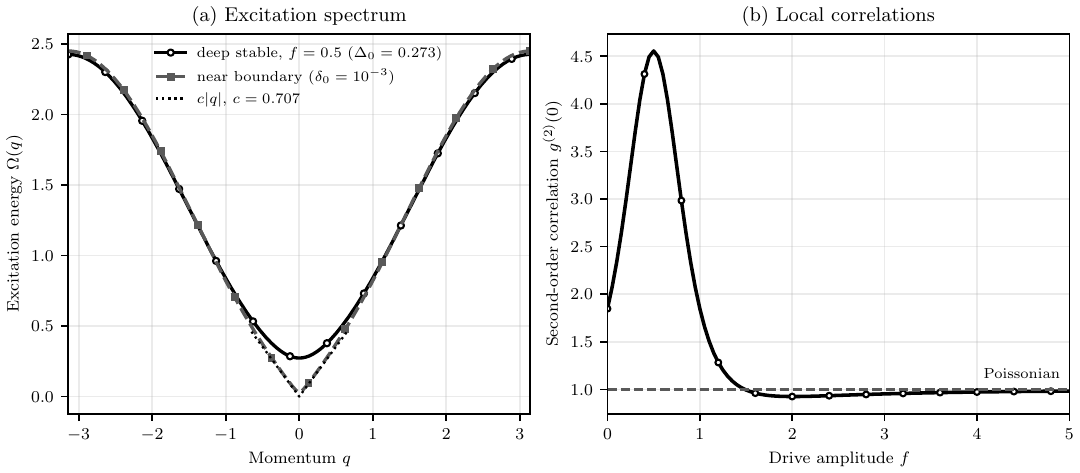}
 \caption{Excitation spectrum and local correlations at $\nu_0=j=1$.
 (a) The solid curve with open-circle markers denotes the gapped, quadratic
 long-wavelength dispersion at the optimized deep-stable point for $f=0.5$;
 the dashed curve with square markers denotes a point having stability margin
 $10^{-3}$, and the dotted curve is the bounding-Hamiltonian boundary
 asymptote $c|q|$ for
 $c=|g|\sqrt{2j}$. (b) The on-site second-order correlation of the stable
 Gaussian family with fixed $(K,g)=(2.0005,0.5)$ as the coherent drive is
 increased. The dashed line marks Poissonian statistics; values above and
 below it indicate bunching and antibunching, respectively.}
 \label{fig:observables-spectrum-correlations}
\end{figure}

Figure~\ref{fig:observables-spectrum-correlations}(a) illustrates the two
long-wavelength regimes of Proposition~\ref{prop:sound-velocity} for
$\nu_0=j=1$.  The deep-stable curve uses the independently optimized
variational parameters at $f=0.5$, $K=1.90722$ and $g=-0.42515$; its
stability margin is approximately $0.09144$ and its gap is
$\Omega(0)\simeq0.27290$, so the dispersion is quadratic at sufficiently
small momentum.  For comparison, the near-boundary curve uses
$K=2.0005$ and $g=0.5$, giving
$\widetilde\nu-j-4g^2=10^{-3}$ and a small residual gap
$\Omega(0)\simeq0.03164$.  Outside the correspondingly narrow quadratic
window it follows the boundary law $\Omega(q)\simeq c|q|$, with

\begin{equation}
 c=|g|\sqrt{2j}=\frac{1}{\sqrt{2}}\simeq0.70711.
 \label{eq:numerical-sound-speed}
\end{equation}

This linear dispersion defines a boundary slope, which may be viewed as an
effective sound velocity of the excitations of the bounding Hamiltonian
$h_1$.
It is not a sound velocity established for the full driven model and does
not mark a phase transition of that model. Whether the full interacting
Hamiltonian has a corresponding low-energy feature must be tested
independently, for example by finite-size scaling or tensor-network methods.
Collective-mode spectroscopy
of Bose--Hubbard systems likewise distinguishes low-energy phase modes from
massive amplitude response~\cite{MenottiTrivedi2008,Endres2012}.

Figure~\ref{fig:observables-spectrum-correlations}(b) holds the same stable
near-boundary pair $(K,g)=(2.0005,0.5)$ fixed and varies only the coherent
drive.  This isolates the displacement-induced change in the Gaussian trial
state.  The moments $m$ and $s$ are evaluated from
Eq.~\eqref{eq:local-moments-tdl}, the filling and number variance from
Eqs.~\eqref{eq:observables-tdl} and~\eqref{eq:number-variance}, and
$g^{(2)}(0)$ from Eq.~\eqref{eq:g2}.  The correlation starts in the bunched
regime, reaches its squeezed-vacuum maximum near $f=g=0.5$, and crosses the
Poissonian value between $f=1.45$ and $1.50$.  At stronger drive it remains
below unity (for example, its minimum is approximately $0.9246$ near $f=2$),
demonstrating drive-induced antibunching and sub-Poissonian fluctuations.
Thus $g^{(2)}(0)<1$ exposes the nonclassical number statistics of this
coherently displaced auxiliary Gaussian state; it is not, by itself, a
prediction for the ground state of the full interacting Hamiltonian.

\section{Numerical Optimization and Tightness of Bounds}
\label{sec:tightness}

Let $|0_B\rangle$ be the normalized ground state of the bounding
Hamiltonian $h_1$, and write the dimensionless Hamiltonian as
$h=T+\sum_{k\in\Lambda}V_k$. Its expectation
$\mu=\langle h\rangle$ in this state is a variational upper bound on the
ground-state energy $E_0$.

\begin{lemma}[Certified Gap for the Full Hamiltonian]
\label{lem:certified-gap}
Let $h$ be the full dimensionless Hamiltonian and $h_1$ the bounding
Hamiltonian of Eq.~\eqref{eq:h1-diagonal}. In finite volume and in the stable
region, the first excited-state energy of $h$ satisfies
\begin{equation}
 E_1(h)\geq E_0(h_1)+\Omega(\mathbf{0}).
 \label{eq:certified-gap}
\end{equation}
\end{lemma}

\begin{proof}[Proof of Lemma~\ref{lem:certified-gap}]
By Proposition~\ref{prop:ly-decomposition},
$h=h_1+\sum_{\mathbf{k}}B_{\mathbf{k}}^\dagger B_{\mathbf{k}}\geq h_1$ in
the quadratic-form sense.  The min--max principle~\cite{ReedSimon4} therefore
gives $E_n(h)\geq E_n(h_1)$ for every $n$.  The diagonal form of $h_1$ has
spectrum
$E_0(h_1)+\sum_{\boldsymbol{\mu}}n_{\boldsymbol{\mu}}
\Omega_{\boldsymbol{\mu}}$, where
$n_{\boldsymbol{\mu}}\in\mathbb Z_{\geq0}$.  Hence
$E_1(h_1)=E_0(h_1)+\min_{\boldsymbol{\mu}}\Omega_{\boldsymbol{\mu}}
=E_0(h_1)+\Omega(\mathbf0)$ by
Proposition~\ref{prop:sound-velocity}, proving the claim.
\end{proof}

We next give an exact and directly implementable expression for the full
variance.  For self-adjoint operators $X$ and $Y$, define
\begin{equation}
 \operatorname{Cov}_{\mathrm{s}}(X,Y)
 =\frac12\left\langle
 (X-\langle X\rangle)(Y-\langle Y\rangle)
 +(Y-\langle Y\rangle)(X-\langle X\rangle)
 \right\rangle .
 \label{eq:symmetrized-covariance}
\end{equation}
With $W=\sum_{k\in\Lambda}V_k$, the exact variance decomposition is
\begin{equation}
 \begin{split}
 \operatorname{Var}(h)
 ={}&\operatorname{Var}(T)
 +2\operatorname{Cov}_{\mathrm{s}}(T,W)\\
 &+\sum_{k,l\in\Lambda}
 \operatorname{Cov}_{\mathrm{s}}(V_k,V_l).
 \end{split}
 \label{eq:variance-h}
\end{equation}
In a translation-invariant state on the periodic lattice, the last term is
\begin{equation}
 M^d\sum_{r\in\Lambda}\operatorname{Cov}_{\mathrm{s}}(V_0,V_r).
 \label{eq:translation-invariant-variance}
\end{equation}
Consequently it cannot in general be replaced by
$M^d\operatorname{Var}(V_0)$: that replacement would require all
inter-site covariances to vanish.

For completeness, define $x_k=\widetilde a_k$ and the translation-invariant
contractions
\begin{equation}
 N_r=\langle x_0^\dagger x_r\rangle,
 \qquad S_r=\langle x_0x_r\rangle,
 \label{eq:variance-contractions}
\end{equation}
which are the inverse discrete Fourier transforms of, respectively,
$(\omega_{\boldsymbol\mu}-\Omega_{\boldsymbol\mu})/(2\Omega_{\boldsymbol\mu})$
and $-g^2/\Omega_{\boldsymbol\mu}$. Thus $N_0=m$ and $S_0=s$.  If
$u=o_1\cdots o_{2p}$ is any even word in the $x_k,x_k^\dagger$, put
\begin{equation}
 {\cal W}(u)=\sum_{\text{pairings }P}\prod_{(i,j)\in P}
 \langle o_i o_j\rangle,
 \qquad {\cal W}(u)=0\quad\hbox{for odd words},
 \label{eq:wick-functional}
\end{equation}
where the four ordered contractions follow from $N_r,S_r$ and
$\langle x_kx_l^\dagger\rangle=\delta_{kl}+N_{k-l}$.  Expanding
\begin{align}
 T&=-\frac{j}{2d}\sum_{\langle k,l\rangle}
 [(x_k^\dagger+\lambda)(x_l+\lambda)+(k\leftrightarrow l)],
 \label{eq:T-shifted}\\
 V_k&=(x_k^\dagger+\lambda)^2(x_k+\lambda)^2
 -\nu_0(x_k^\dagger+\lambda)(x_k+\lambda)
 +f(x_k^\dagger+x_k+2\lambda)
 \label{eq:Vk-shifted}
\end{align}
as $X=\sum_u c_u u$, define the completely explicit bilinear Wick sum
\begin{equation}
 {\cal C}(X,Y)=\frac12\sum_{u,v}c_ud_v
 \{ {\cal W}(uv)+{\cal W}(vu)-2{\cal W}(u){\cal W}(v)\}.
 \label{eq:wick-covariance}
\end{equation}
All words in Eq.~\eqref{eq:wick-covariance} have length at most eight.

\begin{proposition}[Energy Variance in the Bogoliubov Vacuum]
\label{prop:energy-variance}
In the state $|0_B\rangle$, the exact full-Hamiltonian variance is
\begin{equation}
 \operatorname{Var}(h)={\cal C}(T,T)+2{\cal C}(T,W)+{\cal C}(W,W),
 \qquad W=\sum_kV_k.
 \label{eq:energy-variance}
\end{equation}
Equivalently, with $Q=\sum_kB_k^\dagger B_k$,
\begin{equation}
 \operatorname{Var}(h)=\operatorname{Var}(Q)
 ={\cal C}(Q,Q).
 \label{eq:energy-variance-Q}
\end{equation}
Equations~\eqref{eq:variance-contractions}--\eqref{eq:wick-covariance} are
finite lattice sums in $\lambda,m,s,N_r,S_r$ and give the requested exact
expressions for $\operatorname{Var}(T)$,
$\operatorname{Cov}_{\rm s}(T,W)$, and $\operatorname{Var}(W)$.
\end{proposition}

\begin{proof}[Proof of Proposition~\ref{prop:energy-variance}]
The first identity is the algebraic variance decomposition.  The state is
quasi-free, so Wick's theorem evaluates every term exactly as
Eq.~\eqref{eq:wick-covariance}.  Finally $h=h_1+Q$ and
$h_1|0_B\rangle=E_0(h_1)|0_B\rangle$; consequently constants and all mixed
covariances with $h_1$ drop out, giving Eq.~\eqref{eq:energy-variance-Q}.
\end{proof}

\begin{theorem}[Temple's Inequality with Certified Gap]
\label{thm:temple-certified}
Let $\mu=\langle h\rangle$ be the variational upper bound of
Remark~\ref{rem:upper-bound}. If
$\mu<E_0(h_1)+\Omega(\mathbf0)$, then
\begin{equation}
 E_0\geq\mu-\frac{\operatorname{Var}(h)}
 {E_0(h_1)+\Omega(\mathbf0)-\mu}.
 \label{eq:temple-certified}
\end{equation}
Together with Theorem~\ref{thm:lower-bound}, this gives
\begin{equation}
 \max\left\{E^{(d)}_{0,l},\mu-\frac{\operatorname{Var}(h)}
 {E_0(h_1)+\Omega(\mathbf0)-\mu}\right\}
 \leq E_0\leq\mu.
 \label{eq:two-sided-certified}
\end{equation}
\end{theorem}

\begin{proof}[Proof of Theorem~\ref{thm:temple-certified}]
Lemma~\ref{lem:certified-gap} supplies the lower spectral threshold
$E_1^*=E_0(h_1)+\Omega(\mathbf0)$.  The stated condition implies
$\mu<E_1^*\leq E_1(h)$, so Temple's inequality gives
Eq.~\eqref{eq:temple-certified}.  Taking the stronger of this result and
Theorem~\ref{thm:lower-bound}, and using the variational principle for the
upper bound, proves Eq.~\eqref{eq:two-sided-certified}.
\end{proof}

\subsection{Elementary Site and Bond Baselines}
\label{sec:elementary-bounds}

The Lieb--Yamazaki estimate should be judged against two elementary rigorous
lower bounds. Define the one-site ground energy
\begin{equation}
 \varepsilon_1(\eta,f)=\inf\operatorname{spec}
 \left[a^{\dagger2}a^2-\eta a^\dagger a+f(a^\dagger+a)\right]
 \label{eq:one-site-energy}
\end{equation}
and the two-site bond Hamiltonian
\begin{equation}
 h_b=\frac{V_k+V_l}{2d}
 -\frac{j}{2d}(a_k^\dagger a_l+a_l^\dagger a_k),
 \qquad
 V_k=a_k^{\dagger2}a_k^2-\nu_0n_k+f(a_k^\dagger+a_k).
 \label{eq:bond-hamiltonian}
\end{equation}
Then the exact energy density satisfies
\begin{equation}
 \frac{E_0(M)}{M^d}\geq\varepsilon_1(\nu_0+j,f),
 \qquad
 \frac{E_0(M)}{M^d}\geq d\,\inf\operatorname{spec}h_b.
 \label{eq:elementary-lower-bounds}
\end{equation}
Indeed,
$-(a_k^\dagger a_l+a_l^\dagger a_k)\geq-(n_k+n_l)$ follows from
$(a_k-a_l)^\dagger(a_k-a_l)\geq0$; summing bonds gives the site estimate.
For the second estimate, $h=\sum_{\langle k,l\rangle}h_b$ and there are
$dM^d$ bonds, so the Anderson cluster argument~\cite{Anderson1951} applies.

Table~\ref{tab:elementary-bounds} reports a cutoff-converged numerical
comparison in one dimension. The bond bound is substantially sharper than
the Lieb--Yamazaki bound at every listed point, while the site bound can also
be competitive. Thus the value of the present construction lies primarily
in its analytic parameter dependence and its explicitly solvable auxiliary
Gaussian state, not in uniformly optimal energy bounds.
\begin{table}[b]
 \caption{Lower-bound energy densities in $d=1$; larger values are sharper.
 The last column is the converged $M=4$ truncated-Hamiltonian reference.}
 \label{tab:elementary-bounds}
 \begin{ruledtabular}
 \begin{tabular}{ccccc}
 $(\nu_0,j,f)$ & LY & site & bond & $M=4$ ED\\
 \hline
 $(1,1,0)$   & $-2.2500$ & $-2.0000$ & $-1.7321$ & $-1.5988$\\
 $(1,1,0.5)$ & $-2.9622$ & $-2.8930$ & $-2.6415$ & $-2.5987$\\
 $(1,1,2)$   & $-6.5354$ & $-6.5359$ & $-6.3781$ & $-6.3616$\\
 $(1,0.2,5)$ & $-13.4519$& $-13.4032$& $-13.3787$& $-13.3783$\\
 $(1,5,5)$   & $-30.0301$& $-30.1787$& $-29.5555$& $-29.4296$
 \end{tabular}
 \end{ruledtabular}
\end{table}

\subsection{Optimization of the Variational Parameters}

We now optimize the thermodynamic lower-bound energy density of
Corollary~\ref{cor:energy-density} over the admissible decomposition
parameters. For fixed model parameters $(\nu_0,j,f)$, the formal optimum is
the supremum over the open stable set,
\begin{equation}
 e_l^*=\sup_{\substack{K>\frac12(1+\nu_0+j+4g^2)\\ |g|\leq2}}
 e^{(1)}_{0,l}(K,g).
 \label{eq:formal-optimal-lower-bound}
\end{equation}
For numerical work we keep a distance $\epsilon$ from the boundary and define
\begin{equation}
 e_l^*(\epsilon)=
 \max_{\substack{(K,g)\in\mathcal{S}_\epsilon\\ |g|\leq2}}
 e^{(1)}_{0,l}(K,g),
 \qquad
 \mathcal{S}_\epsilon=\left\{(K,g):
 K\geq\frac12(1+\nu_0+j+4g^2)+\epsilon\right\},
 \label{eq:optimal-lower-bound}
\end{equation}
and denote an optimizing pair by $(K_l^*,g_l^*)$. The parameters $K$ and $g$
belong to the Lieb--Yamazaki decomposition; this optimization is over those
variational parameters, not a limit such as $g\to0$. We use
$\epsilon=10^{-4}$ in the numerical calculations. At regular boundary points
the energy density has a finite continuous limit, so this regulator is not
part of the formal supremum; it stabilizes the numerical quadrature. In one
dimension, however, individual Bogoliubov moments and the same-pair upper
estimate can diverge or depend logarithmically on $\epsilon$ near a gapless
optimizer. We distinguish two
upper-bound energy-density protocols. The upper-bound energy density evaluated
at the lower-bound optimizer is
\begin{equation}
 e_u^{\text{at-lower}}\equiv e_u(K_l^*,g_l^*),
 \label{eq:upper-at-lower}
\end{equation}
whereas independent upper-bound energy-density optimization gives
\begin{equation}
 e_u^{\min}\equiv
 \min_{(K,g)\in\mathcal{S}_\epsilon}e_u(K,g).
 \label{eq:upper-min}
\end{equation}
For brevity, figures label $e_l^*(\epsilon)$ as $e_l^*$. The associated
$e_u^{\text{at-lower}}$ is explicitly regulator dependent, whereas
$e_u^{\min}$ is the primary upper estimate used in the regime maps. All
numerical multistart searches, including the independent
minimization in Eq.~\eqref{eq:upper-min}, are restricted to $|g|\leq2$;
their reported best points are not proofs of global extrema. Only the
same-pair gap obeys the decomposition identity
\begin{equation}
 \Delta_{\text{at-lower}}
 =e_u^{\text{at-lower}}-e_l^*
 =\left\langle B_0^\dagger B_0\right\rangle_{(K_l^*,g_l^*)}
 \geq0,
 \label{eq:bound-gap}
\end{equation}
where the equality follows from the positive-operator decomposition and the
fact that $|0_B\rangle$ is the ground state of $h_1$. For the independently
optimized bracket one has only $e_u^{\min}-e_l^*\geq0$ by the variational
and lower-bound principles.

\begin{figure}[p]
 \centering
 \includegraphics[width=\columnwidth]{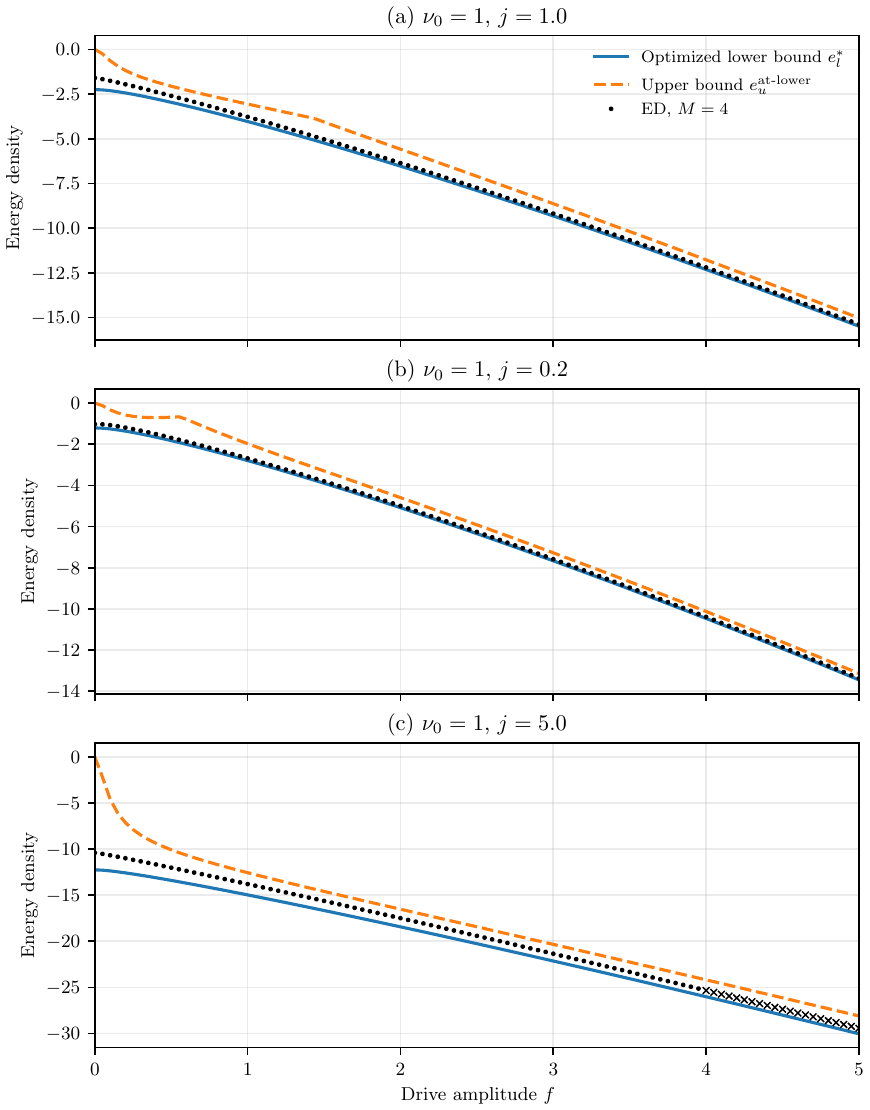}
 \caption{Optimized thermodynamic lower-bound energy density $e_l^*$ and
 variational upper-bound energy density $e_u^{\text{at-lower}}$ evaluated at the same lower-bound
 optimizer versus drive amplitude over $0\leq f\leq5$ for (a) $j=1.0$,
 (b) $j=0.2$, and (c) $j=5.0$, at fixed $\nu_0=1$. Small black dots are
 converged eigenvalues of the $n_{\max}=10$ truncated periodic chain with
 $M=4$ outside the mixed strong regime; black crosses for $f\geq4$, $j\geq4$
 use adaptive $n_{\max}=14$--18 with successive energy-density changes below
 $10^{-4}$ (see Fig.~\ref{fig:nmax-convergence}). These finite-chain markers
 are numerical references, not TDL energy densities. The numerical step is
 $\Delta f=0.05$.}
 \label{fig:bounds_vs_f}
\end{figure}

\begin{figure}[p]
 \centering
 \includegraphics[width=\columnwidth]{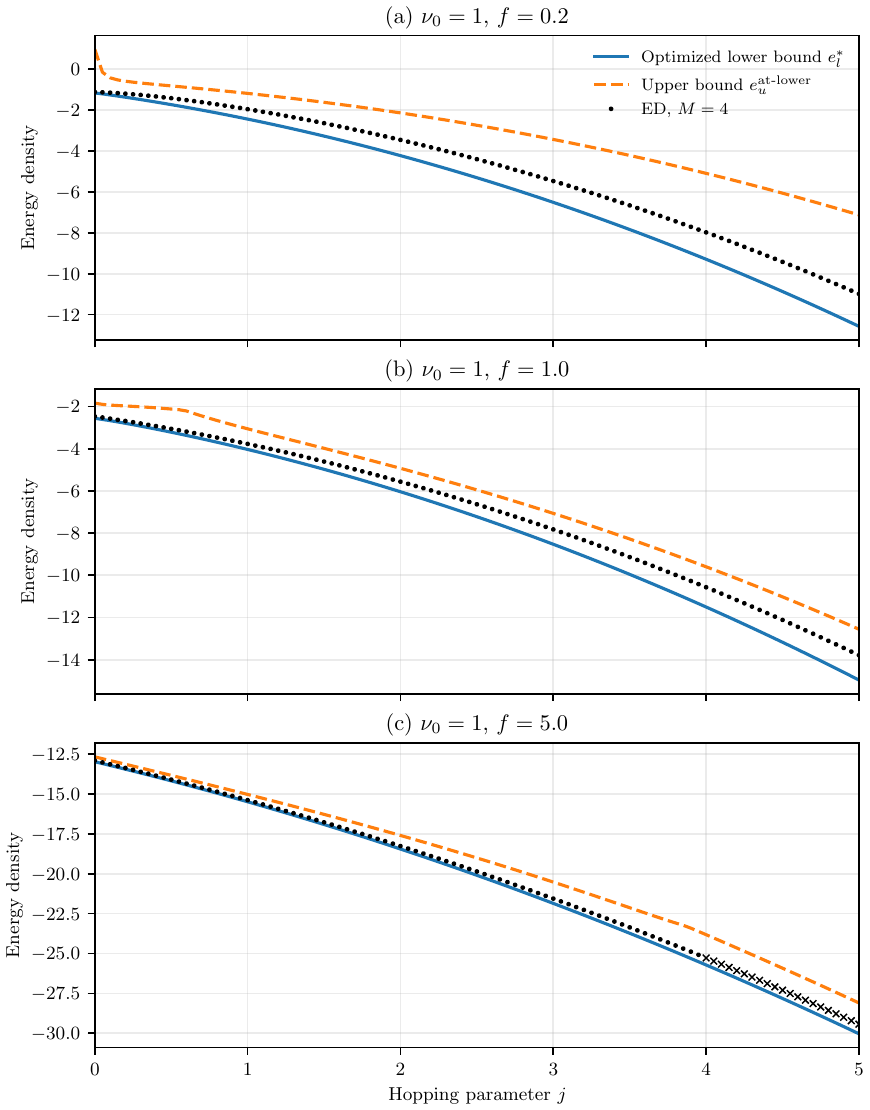}
 \caption{Optimized thermodynamic lower-bound energy density $e_l^*$ and
 variational upper-bound energy density $e_u^{\text{at-lower}}$ evaluated at the same lower-bound
 optimizer versus hopping parameter over $0\leq j\leq5$ for (a) $f=0.2$,
 (b) $f=1.0$, and (c) $f=5.0$, at fixed $\nu_0=1$. Small black dots are
 converged eigenvalues of the $n_{\max}=10$ truncated periodic chain with
 $M=4$ outside the mixed strong regime; black crosses for $f\geq4$, $j\geq4$
 use adaptive $n_{\max}=14$--16 with successive energy-density changes below
 $10^{-4}$ (see Fig.~\ref{fig:nmax-convergence}). These finite-chain markers
 are numerical references, not TDL energy densities. The numerical step is
 $\Delta j=0.05$.}
 \label{fig:bounds_vs_j}
\end{figure}

\begin{figure}[p]
 \centering
 \includegraphics[width=\columnwidth]{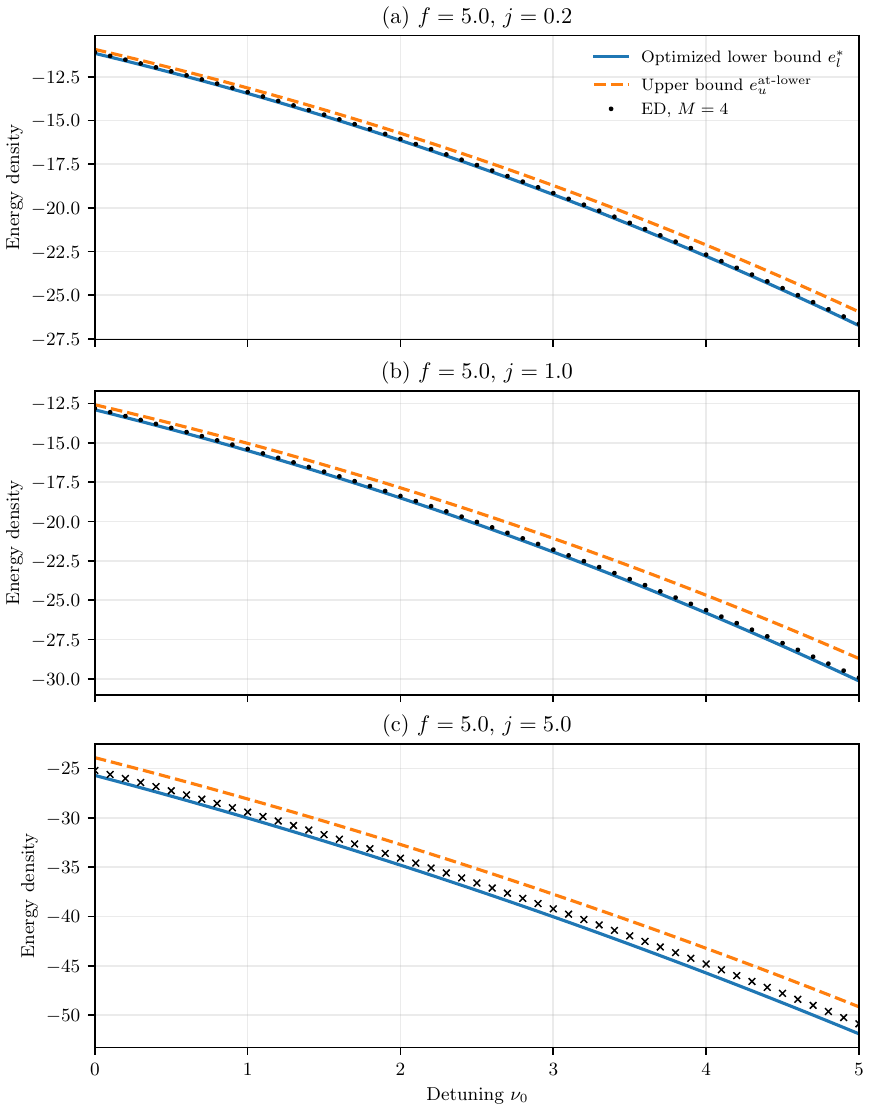}
 \caption{Optimized thermodynamic lower-bound energy density $e_l^*$ and
 variational upper-bound energy density $e_u^{\text{at-lower}}$ evaluated at the same lower-bound
 optimizer versus detuning over $0\leq\nu_0\leq5$, beginning at exact resonance,
 for (a) $j=0.2$, (b) $j=1.0$, and (c) $j=5.0$, at fixed strong drive $f=5$.
 Small black dots are converged eigenvalues of the $n_{\max}=10$ truncated
 periodic chain with $M=4$ for $j=0.2$ and $1.0$; black crosses for $j=5$
 use adaptive $n_{\max}=14$--18 with successive energy-density changes below
 $10^{-4}$ (see Fig.~\ref{fig:nmax-convergence}). These finite-chain markers
 are numerical references, not TDL energy densities. The numerical step is
 $\Delta\nu_0=0.1$.}
 \label{fig:bounds_vs_nu0}
\end{figure}

\begin{figure}[p]
 \centering
 \includegraphics[width=\columnwidth]{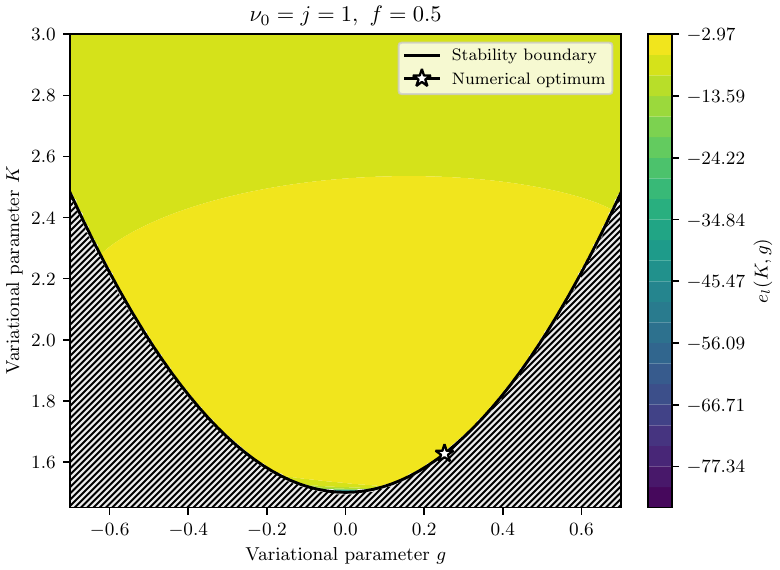}
 \caption{Thermodynamic lower-bound energy-density landscape
 $e^{(1)}_{0,l}(K,g)$ for
 $\nu_0=j=1$ and $f=0.5$. The hatched region is outside the stability
 constraint regularized by $\epsilon=10^{-4}$, and the star indicates the
 numerical maximum. The optimization uses a nine-start L-BFGS-B search over
 $|g|\leq2$ inside the stable region.}
 \label{fig:optimization_landscape}
\end{figure}

Figure~\ref{fig:bounds_vs_f} compares the optimized thermodynamic lower-bound
energy density $e_l^*$ with the upper-bound energy density
$e_u^{\text{at-lower}}$ throughout $0\leq f\leq5$ at fixed
$\nu_0=1$ for the reference
hopping $j=1$, weak hopping $j=0.2$, and strong hopping $j=5$. Numerically
converged eigenvalues of the $n_{\max}$-truncated periodic $M=4$ chain are
included in every panel. We retain $n_{\max}=10$ outside the mixed strong
regime; every marker with $f\geq4$ and $j\geq4$ was recomputed adaptively at
$n_{\max}=12,14,16,18$ until consecutive energy densities differed by less
than $10^{-4}$. These points are exact for their finite-dimensional truncated
Hamiltonians and serve as numerical reference values, rather than exact
thermodynamic energy densities. Exact diagonalization is
restricted by this rapid Hilbert-space growth; matrix-product-state and DMRG
methods provide the standard scalable alternative in one dimension
~\cite{White1992,Schollwock2011,Orus2014,HauschildPollmann2018,Paeckel2019}.
In panel (a), at $f=0.5$, the
optimization gives $K_l^*\simeq1.6268$, $g_l^*\simeq0.2517$,
$e_l^*\simeq-2.9623$, and
$e_u^{\text{at-lower}}\simeq-2.0378$, so the relative same-pair gap
$\Delta_{\text{at-lower}}/|e_l^*|$ is about $31.2\%$. This number is
regulator dependent near the stability boundary. Unless independent
minimization is stated explicitly, all gaps discussed for
Figs.~\ref{fig:bounds_vs_f}--\ref{fig:bounds_vs_nu0} use this same-pair
protocol. Comparing the panels shows that weak
hopping produces the narrowest relative envelope at the upper end of the
scan: at $f=5$ the relative gaps are approximately $3.0\%$, $2.3\%$, and
$6.4\%$ for $j=1$, $0.2$, and $5$, respectively. Thus all three envelopes
tighten with increasing drive, although the approach is slowest in the
strong-hopping case. The color map in Fig.~\ref{fig:optimization_landscape} shows
the lower-bound landscape at $f=0.5$ and marks its numerically found maximum
in the stable region.

The larger cutoff is essential in the mixed strong regime. At $j=f=5$,
increasing $n_{\max}$ from $6$ to $10$ lowers the truncated-chain energy
density from approximately $-27.4099$ to $-29.4181$; the adaptive calculation
converges at $n_{\max}=16$ to approximately $-29.4296$. The converged value
lies between the thermodynamic lower and upper curves. Thus the earlier
points above the upper curve were a local-Hilbert-space truncation artifact.
The reduced cutoff error makes placement inside the thermodynamic envelope a
reliable numerical diagnostic at the plotted resolution.

Figure~\ref{fig:bounds_vs_j} gives the complementary scan in the hopping
parameter at fixed drive. For weak drive, $f=0.2$, the absolute gap has a
shallow minimum of approximately $0.78$ near $j=0.2$ and then broadens
substantially to about $5.45$ at $j=5$. At strong drive, $f=5$, it grows much
more gently, from approximately $0.31$ at $j=0$ to $1.93$ at $j=5$; relative
to $|e_l^*|$, the corresponding endpoint gaps are about $2.4\%$ and $6.4\%$.
At every sampled point the finite-chain ED marker lies between the two
thermodynamic curves.

\clearpage

Figure~\ref{fig:bounds_vs_nu0} starts at exact resonance, $\nu_0=0$, and
tracks the bounds with increasing detuning at fixed strong drive $f=5$. At
resonance the absolute gaps are approximately $0.228$, $0.320$, and $1.800$
for $j=0.2$, $1$, and $5$, respectively. The corresponding stability margins
$\min_\mu(\omega_\mu-2g^2)$ are approximately $0.7251$, $0.3108$, and
$2\times10^{-4}$, so every optimum is stable; the strong-hopping solution
lies close to the regularized stability boundary. The absolute gaps increase
monotonically over the plotted detuning interval, reaching approximately
$0.786$, $1.423$, and $2.739$ at $\nu_0=5$. All finite-chain ED points again
lie between the two thermodynamic curves throughout the scan.

\begin{figure}[p]
 \centering
 \includegraphics[width=\columnwidth]{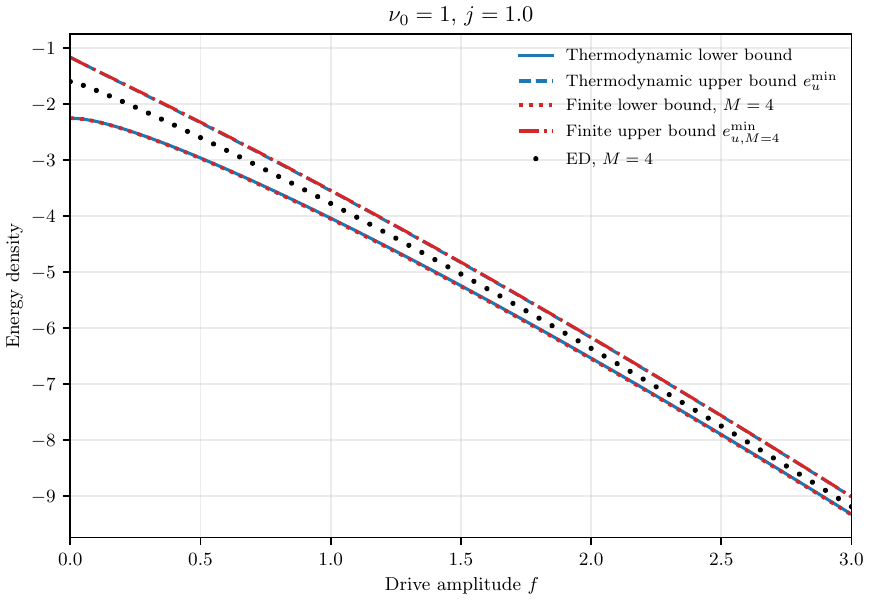}
 \caption{Finite-size comparison at $\nu_0=1$, $j=1$, and
 $0\leq f\leq3$. Blue solid and dashed curves are the optimized
 thermodynamic Lieb--Yamazaki lower-bound energy density $e_l^*$ and
 independently minimized variational upper-bound energy density $e_u^{\min}$. Red dotted
 and dash-dotted curves are their finite-volume $M=4$ counterparts. Each
 upper curve is independently minimized over $K$ and $g$ inside the stable
 region, while each lower curve is independently maximized there.
 Black points are converged eigenvalues of the $n_{\max}=12$ truncated
 Hamiltonian for $M=4$; their changes from $n_{\max}=10$ are below
 $1.1\times10^{-8}$, well within the $10^{-4}$ criterion of
 Fig.~\ref{fig:nmax-convergence}. These finite-chain markers are numerical
 references, not TDL energy densities.
 The finite-volume pair isolates the ansatz error at fixed size, while its
 displacement from the blue curves displays finite-size effects.}
 \label{fig:finite-volume-bounds}
\end{figure}

Figure~\ref{fig:finite-volume-bounds} separates finite-size effects from
the thermodynamic approximation at $\nu_0=j=1$. The finite-volume
Lieb--Yamazaki lower bound uses the four discrete momenta of the $M=4$ ring
and rigorously bounds its exact ground-state energy from below. Evaluating
the finite-volume variational energy and minimizing it independently over
$K$ and $g$ in the stable region supplies the corresponding optimized upper
bound. The thermodynamic upper curve is minimized in the same way. The
separation of each lower--upper pair is therefore a direct measure of the
best bracket obtained from the Bogoliubov ansatz for that system size,
independently of the thermodynamic limit. Numerically, every recomputed
$n_{\max}=12$ ED point lies inside this finite-volume bracket over
$0\leq f\leq3$; the smallest observed distances from the lower and upper
curves are approximately $0.144$ and $0.179$, respectively. The cutoff ED
values approximate the untruncated finite-volume energy to which the rigorous
bounds apply. The finite-volume and thermodynamic upper bounds differ by at
most approximately $0.0100$ over the scan and their curves are nearly
superposed. The same is true of the lower bounds, whose maximum absolute
difference is approximately $0.0120$. Thus finite-size corrections to both
optimized bounds are small for $M=4$ in this parameter range, even though the
finite-volume curves are the ones that rigorously bracket the finite system.

The four curves use four separate optimizing pairs $(K,g)$ at every drive.
For example, Table~\ref{tab:four-optimizers} gives the values at $f=0.2$.
The lower-bound pairs maximize the corresponding Lieb--Yamazaki functional,
whereas the upper-bound pairs independently minimize the variational energy;
all four optimizations are restricted to the stable region.

\begin{table}[ht]
 \caption{The four independently optimized stable parameter pairs at
 $f=0.2$, $\nu_0=1$, and $j=1$.}
 \label{tab:four-optimizers}
 \begin{ruledtabular}
 \begin{tabular}{lcc}
 Bound & $K^*$ & $g^*$ \\
 \hline
 Thermodynamic lower & $1.540123$ & $0.141461$ \\
 Thermodynamic upper & $1.753143$ & $-0.341058$ \\
 Finite-volume lower, $M=4$ & $1.539911$ & $0.141042$ \\
 Finite-volume upper, $M=4$ & $1.745619$ & $-0.323582$ \\
 \end{tabular}
 \end{ruledtabular}
\end{table}

\clearpage

\subsection{Numerical Assessment of the Temple Estimate}

For a translation-invariant Bogoliubov vacuum, the decomposition gives
\begin{equation}
 \mu-E_0(h_1)=\langle Q\rangle
 =M^d\langle B_0^\dagger B_0\rangle=M^d\Delta_M .
 \label{eq:temple-extensive-excess}
\end{equation}
The Temple applicability condition is therefore
\begin{equation}
 M^d\Delta_M<\Omega(\mathbf0).
 \label{eq:temple-size-condition}
\end{equation}
Unless $\Delta_M$ vanishes with volume, its left-hand side is extensive while
the certified one-particle gap is intensive. Hence a fixed trial family can
satisfy the test only below a parameter-dependent size threshold; closeness
to the stability boundary makes the condition still more restrictive because
$\Omega(\mathbf0)$ closes. At a given point, the auxiliary gap would have to
exceed $M^d\Delta_M$; equivalently, if the present test fails, it must increase
by a factor greater than $M^d\Delta_M/\Omega(\mathbf0)$ while the excess is
held fixed. This is a necessary threshold, not evidence that any particular
extension of the ansatz will achieve it. A finite cutoff imposed directly on the Gaussian
state can also spoil this identity, so numerical tests must evaluate the
Gaussian moments consistently, for example through the Wick formulas above.

This scaling observation replaces a universal negative verdict based on a
small cutoff scan. At a particular finite size the condition may pass or
fail, and a different admissible decomposition may improve the threshold,
but neither outcome by itself diagnoses the spectrum of the full model.
Equation~\eqref{eq:temple-certified} remains a valid conditional finite-volume
refinement; it does not yield a thermodynamic energy-density correction for
a nonzero limiting $\Delta_M$.

\subsection{Regime Maps and Limitations of the Ansatz}

\begin{figure}[p]
 \centering
 \includegraphics[width=\columnwidth]{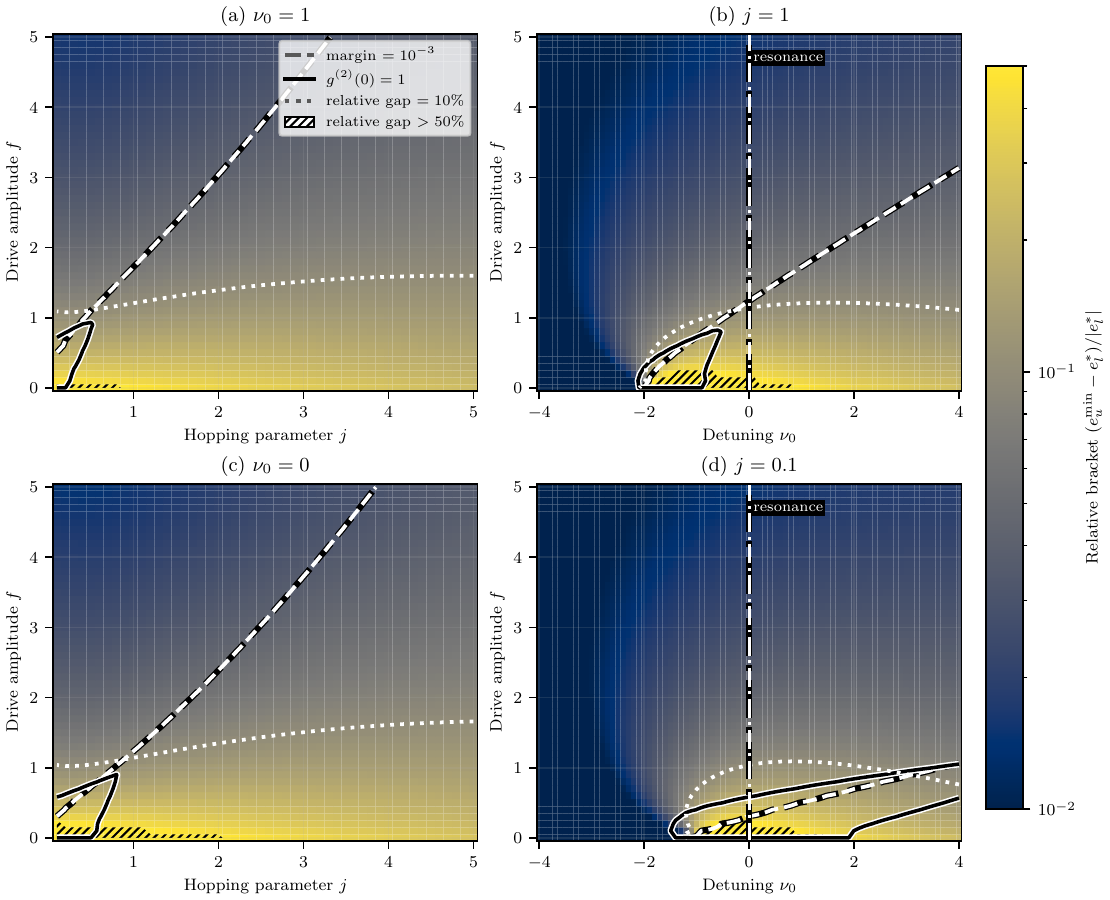}
 \caption{Regime maps of the optimized thermodynamic bounds on (a) the
 $(j,f)$ plane at $\nu_0=1$, (b) the $(\nu_0,f)$ plane at $j=1$,
 (c) the $(j,f)$ plane at exact resonance $\nu_0=0$, and (d) the
 $(\nu_0,f)$ plane at weak hopping $j=0.1$.  Both detuning panels cover
 $-4\leq\nu_0\leq4$.
 Color, with logarithmic color normalization, represents the relative bracket width
 $(e_u^{\min}-e_l^*)/|e_l^*|$, where $e_u^{\min}$ is independently minimized,
 clipped at $50\%$. Hatched regions have a relative bracket exceeding
 $50\%$. Thick dashed and dotted contours denote, respectively, the stability
 margin $\min_\mu(\omega_\mu-2g^2)=10^{-3}$ and a relative gap of $10\%$.
 The haloed solid contour denotes $g^{(2)}(0)=1$;
 the white dash-dotted lines in the detuning panels mark exact resonance.
 Every
 displayed optimum is stable, so no hatched unstable region is present.}
 \label{fig:phase-diagram}
\end{figure}

The four regime maps in Fig.~\ref{fig:phase-diagram} summarize where the
optimized Bogoliubov construction is quantitatively tight.  Panels (a) and
(c) compare the $(j,f)$ plane at detunings $\nu_0=1$ and $0$, while panels
(b) and (d) compare the $(\nu_0,f)$ plane over $-4\leq\nu_0\leq4$ at
hoppings $j=1$ and $0.1$.
The relative gap is below $10\%$ throughout all four sampled planes by
$f=2$; the numerical contour is
already reached between approximately $f=1.1$ and $1.6$, depending on the
remaining parameter.  Consequently the especially conservative statement
that the bounds are narrow for strong drive $f\gtrsim3$ and moderate hopping
$j\lesssim2$ is satisfied, as is the corresponding statement for
$\nu_0\geq1$ and $f\geq2$.  The map in fact shows that the latter precision
extends over the full sampled hopping and detuning ranges at $f\geq2$.

The red dashed contour marks a stability margin of $10^{-3}$.  All optimized
points shown are inside the stable region, as required by the construction;
the contour identifies the narrow strip in which the solution approaches
its boundary.  At large hopping, in particular near $j=5$, the lower-bound
optimizer remains close to this contour over the displayed drive interval,
which helps explain the comparatively wider bracket there.  Crossing the
actual boundary would make the Bogoliubov frequencies imaginary and the
transformation physically inadmissible.

The black contours in Fig.~\ref{fig:phase-diagram} mark
$g^{(2)}(0)=1$ for the Gaussian state selected by the lower-bound optimum.
In panels (a) and (c) they separate a bunched region at weak hopping and
weak drive from the antibunched region generated by coherent driving; exact
resonance enlarges the bunched region.  At weak hopping $j=0.1$, panel (d)
similarly exhibits a pronounced bunched tongue at small drive.  Extending
the detuning range to negative values also reveals a bunched pocket in panel
(b), centered on weak drive and negative detuning, which was absent from the
previously displayed nonnegative-detuning window.  These are correlation
crossovers, not phase transitions.

As a reading guide, the upper portions of all four panels ($f\gtrsim2$) are
the narrow-bracket regime, below the dotted $10\%$ contour. The low-drive
parts contain the widest brackets and all crossings of the haloed
$g^{(2)}(0)=1$ contour, while the dashed stability-margin contour identifies
where the optimized auxiliary Hamiltonian approaches its gapless boundary.

The map should not be read as a conventional equilibrium phase diagram.
In particular, this method does not reproduce the Mott lobes of the undriven
Bose--Hubbard model.  The coherent term $f(a_k^\dagger+a_k)$ explicitly
breaks the number-conserving $U(1)$ symmetry, while the uniform displacement
$\lambda$ builds the ansatz around a coherent state.  Thus every $f>0$
induces some coherence and removes the sharp symmetry-based Mott transition,
although smooth remnants and crossovers may remain.

The Gaussian Bogoliubov vacuum also cannot capture all strong non-Gaussian
correlations that may become important at very large hopping or close to the
stability boundary.  The widening lower--upper bracket in these regions is
therefore a useful diagnostic that more elaborate correlated states or
methods such as DMRG may be needed~\cite{Schollwock2011,Paeckel2019}.  It is
not a failure of rigor: the
Lieb--Yamazaki lower bound remains a strict lower estimate throughout the
stable region, while the independently minimized Gaussian expectation is a
variational upper estimate.  Figures~\ref{fig:bounds_vs_f}--
\ref{fig:bounds_vs_nu0} likewise place the $M=4$ ED data inside the reported
brackets.  For example, evaluating the upper energy density at the lower-bound
optimum gives, at $f=0.5$, $j=\nu_0=1$,
$\Delta_{\text{at-lower}}/|e_l^*|\simeq31.2\%$ at $\epsilon=10^{-4}$.
This same-pair value is regulator dependent near the stability boundary.
Independent upper
minimization used in the present map gives
$(e_u^{\min}-e_l^*)/|e_l^*|\simeq21.2\%$. The reduction measures how rigidly
the Gaussian state tied to the lower-bound optimum brackets the energy versus
what the same Gaussian family can achieve under a separate variational
minimization; the residual width diagnoses limitations of that ansatz rather
than a numerical or mathematical inconsistency.

\subsection{Convergence with the Local Occupation Cutoff}

\begin{figure}[p]
 \centering
 \includegraphics[width=\columnwidth]{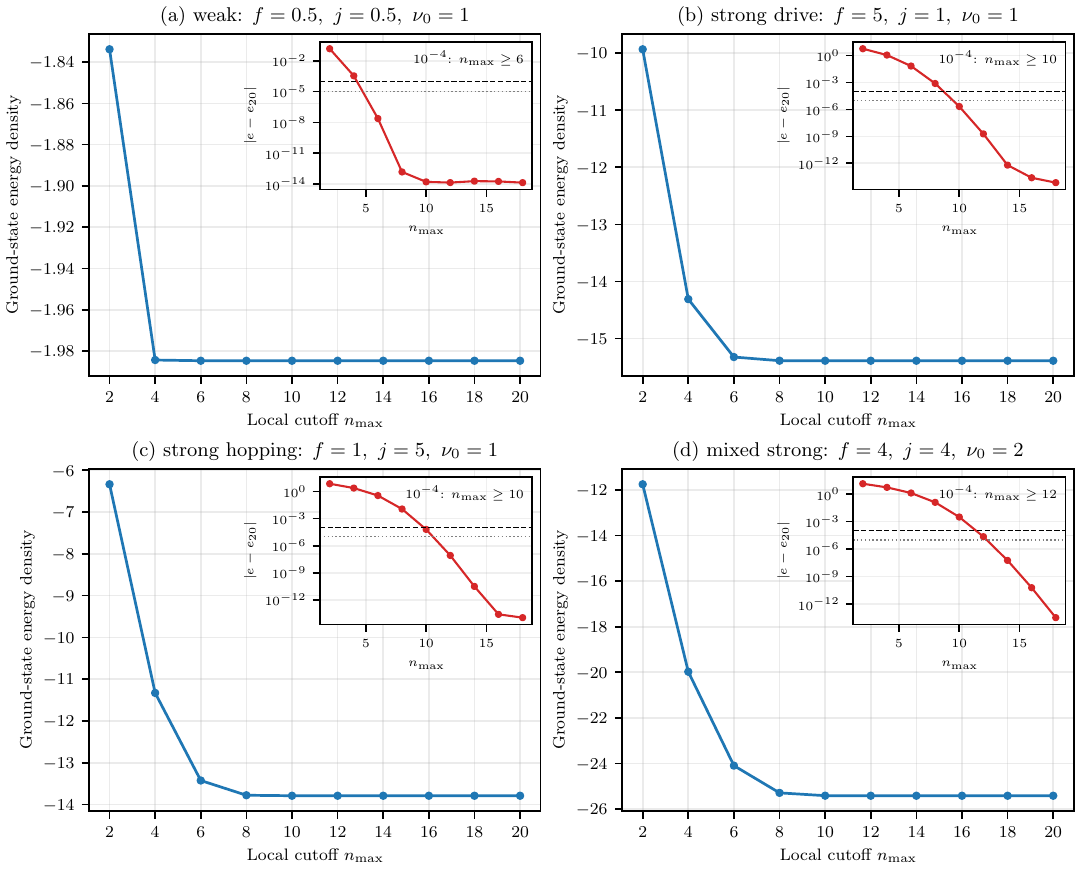}
 \caption{Convergence of the $M=4$ ED ground-state energy density with the
 local occupation cutoff.  The panels show (a) the weak case
 $(f,j,\nu_0)=(0.5,0.5,1)$, (b) strong drive $(5,1,1)$, (c) strong hopping
 $(1,5,1)$, and (d) the mixed strong case $(4,4,2)$.  Insets display
 $|e(n_{\max})-e(20)|$ on a logarithmic scale; dashed and dotted horizontal
 lines mark $10^{-4}$ and $10^{-5}$, respectively.}
 \label{fig:nmax-convergence}
\end{figure}

We finally tested the local Fock-space truncation used in the finite-chain
ED data.  Figure~\ref{fig:nmax-convergence} follows the $M=4$ ground-state
energy density through $n_{\max}=20$ in four representative regimes.  Using
$e(20)$ as the reference, the absolute error remains below $10^{-4}$ at all
subsequent sampled cutoffs from $n_{\max}=6$, 10, 10, and 12 in cases A--D,
respectively.  The corresponding cutoffs for the stricter $10^{-5}$ test are
6, 10, 12, and 14.  Thus the default $n_{\max}=10$ is amply converged for the
weak, strong-drive, and strong-hopping examples.  It is not uniformly
adequate in the mixed strong regime: for case D its error at $n_{\max}=10$
is $3.13\times10^{-3}$, and $n_{\max}\geq12$ (or 14 for $10^{-5}$ accuracy)
should be used there.
The criterion used throughout is therefore an absolute energy-density change
below $10^{-4}$ between consecutive adaptive cutoffs, supplemented by the
explicit convergence study against $e(20)$; no unverified relative
$10^{-6}$ criterion is assumed.

\clearpage

\section{Conclusions}

A family of rigorous lower bounds has been derived for the ground-state
energy of the coherently driven Bose--Hubbard model on periodic hypercubic
lattices.  By the positive-operator decomposition, followed by displacement
and Bogoliubov diagonalization, the lower-bound energy density
$e^{(d)}_{0,l}$ is obtained and its
thermodynamic limit is established for every $d\geq1$.  By the same construction there is produced the
continuous band $\Omega(\mathbf q)$ of
Proposition~\ref{prop:continuous-spectrum}, including its gap, effective
mass, and boundary slope (an effective sound velocity of the excitations of
the bounding Hamiltonian $h_1$, not of the full driven model).  Within the ground
state of the bounding Hamiltonian, the order parameter, filling, energy
densities, number variance, $g^{(2)}(0)$, and anomalous correlators are all
explicit; in
particular, the formulas identify the regime of drive-induced antibunching
and sub-Poissonian fluctuations.

Finite-volume optimization for the $M=4$ ring changes either optimized bound
from its thermodynamic-limit counterpart by at most approximately
$1.2\times10^{-2}$ over the reported scan.  The dominant width of the energy
bracket therefore reflects the limitations of the Gaussian ansatz rather
than finite-size effects.  The independently optimized upper bound and the
regime maps further locate where this bracket widens and trace the
$g^{(2)}(0)=1$ crossover curves separating antibunched and bunched
correlations.  Convergence of the ED markers is parameter dependent:
$n_{\max}=10$ meets the $10^{-4}$ criterion in the weak and axis-strong
regimes, whereas mixed strong regimes require adaptive local cutoffs starting
at $n_{\max}=12$ and extending through $n_{\max}=18$ when necessary.

All reported observables, including $g^{(2)}(0)$, the excitation spectrum
$\Omega(\mathbf q)$, and the boundary slope $c$, are rigorous properties of
the ground state $|0_B\rangle$ of the auxiliary bounding Hamiltonian $h_1$.
They are not established observables or ground-state properties of the full
driven Hamiltonian $h$. Establishing whether the auxiliary gapless boundary
has a counterpart in the full model requires independent finite-size scaling
or tensor-network calculations such as DMRG or iTEBD.

The accuracy attainable with this quadratic reference state is also delimited
by the numerical results.  For the representative parameters
$\nu_0=j=1$ and $f=0.5$, the same-pair separation
$e_u^{\text{at-lower}}-e_l^*$ is about $31.2\%$ of the lower-bound
magnitude at $\epsilon=10^{-4}$, although this same-pair number is regulator
dependent near a gapless stability boundary; the independently optimized
upper bound is the more robust numerical comparator. Lemma~\ref{lem:certified-gap}
makes the Temple test rigorous in finite volume, but its applicability
condition reads $M^d\Delta_M<\Omega(\mathbf0)$. Thus an extensive trial-state
energy excess eventually overwhelms the intensive certified gap. Temple's
inequality remains a conditional finite-size refinement, not a general
thermodynamic correction or a test of whether the ansatz controls the full
spectrum.

The elementary site and Anderson bond bounds added in
Sec.~\ref{sec:elementary-bounds} are sharper at several representative
points, which sets a clear benchmark for the present analytic construction.
Natural extensions include finite-temperature bounds, non-hypercubic or
inhomogeneous geometries, and bounding states that incorporate non-Gaussian
or higher-order correlations. The exactly soluble $h_1$ is thus retained as
a Gaussian reference \textit{par excellence}, rather than as a surrogate for the full
interacting theory. Enlarging the positive-operator family or
combining it with systematically improved variational states may increase
the certified excitation threshold and reduce the energy variance, the two
quantities that govern whether the finite-volume Temple condition can hold.

\begin{acknowledgments}
OpenAI Codex CLI and Qwen were used as language and coding assistance tools
during the preparation and revision of this manuscript. All suggestions and
generated material were critically reviewed by the authors, who bear full
responsibility for the content of the work.

M.J. would like to dedicate this work to the memory of his Mother.
\end{acknowledgments}

\bibliographystyle{aipnum4-2}
\bibliography{references}

\end{document}